\documentclass{aa} 
\usepackage{aadefs}
\usepackage{graphicx}
\usepackage{txfonts}
\usepackage{amsmath,amsfonts,amssymb} 
\usepackage{natbib}
\usepackage{here}
\usepackage{comment}
\usepackage{longtable}
\usepackage{caption}
\bibpunct{(}{)}{;}{a}{}{,}
\usepackage[colorlinks=true,urlcolor=blue,linkcolor=blue,citecolor=blue]{hyperref}
\usepackage[normalem]{ulem}
\usepackage{color}
\usepackage[normalem]{ulem}
\usepackage{xcolor}
\usepackage{cuted}

\usepackage{xcolor}

\begin{document} 
        
\title{Variations in Volatile-Driven Activity of Comet C/2017 K2 (PanSTARRS) Revealed by Long-Term Multi-Wavelength Observations}

\authorrunning{Hmiddouch et al. 2025}
\titlerunning{Comet C/2017 K2 (PanSTARRS)}
        
\author{
 S. Hmiddouch\inst{1,2}$^\ast$, E. Jehin\inst{1}, M. Lippi\inst{3}, M. Vander Donckt\inst{1}, K. Aravind\inst{1}, D. Hutsemékers\inst{1}, J. Manfroid\inst{1}, A. Jabiri\inst{2},\\
 Y. Moulane\inst{4}, \and Z. Benkhaldoun\inst{2}
 }

\institute{
Space sciences, Technologies \& Astrophysics Research (STAR) Institute, University of Liège, Liège, Belgium
\and
Cadi Ayyad University (UCA), Oukaimeden Observatory (OUCA), Faculté des Sciences Semlalia (FSSM),\\
High Energy Physics, Astrophysics and Geoscience Laboratory (LPHEAG), Marrakech, Morocco
\and
INAF – Osservatorio Astrofisico di Arcetri – Largo Enrico Fermi, 5, 50125 Firenze, Italy
\and
School of Applied and Engineering Physics, Mohammed VI Polytechnic University, Ben Guerir 43150, Morocco\\
$^\ast$\email{\color{blue}shmiddouch@uliege.be}
}

\date{Received/accepted}
 
  \abstract
   {By conducting a comprehensive study of comets across a broad range of heliocentric distances, we can improve our understanding of the physical mechanisms that trigger their activity at various distances from the Sun. At the same time, we can identify possible differences in the composition of these outer solar system bodies belonging to various dynamical groups. C/2017 K2 (PANSTARRS) is a Dynamically New Oort Cloud comet (DNC) that has exhibited activity at an extremely large heliocentric distance of 23.75 au, was found to have a CO-rich coma at 6.72 au, and attracted the interest of the community to become a bright DNC at perihelion.}
  {We aim to study through photometry and spectroscopy the activity evolution and chemical composition of C/2017 K2 over a long-term monitoring from October 2017 (r$_h$ = 15.18 au) pre-perihelion to April 2025 (r$_h$ = 8.46 au) post-perihelion.}
   {We used both TRAPPIST telescopes to monitor the activity with broad-band and cometary narrow-band filters. We produced an 8-year light curve and colors from the broad-band images and computed the activity slopes. We derived the production rates of the daughter species, OH, NH, CN, C$_3$, and C$_2$, using a Haser model as well as the dust proxy parameter A(0)f$\rho$. We used CRIRES$^+$, the high-resolution infrared echelle spectrometer of the ESO VLT, and UVES, its high-resolution ultraviolet-visual echelle spectrograph, to study simultaneously the parent and daughter species, at three different epochs from May to September 2022, when the comet crossed the water sublimation region.}
   {C/2017 K2 (PanSTARRS) light curve shows a complex evolution of its brightness, with several slopes and a plateau around perihelion, revealing the onset and competition of various species. The photometric analysis shows constant coma colors across the whole heliocentric range, indicating similar properties of the dust grains released during the survey, and in agreement with the colors of other active long-period comets. The production rates indicate a typical C$_2$/CN composition and a high dust-to-gas ratio. The analysis of the oxygen forbidden lines shows the transition between CO and CO$_2$ dominating the comet's activity to the onset of the water sublimation below 3.0 au. Molecular abundance analysis from the IR spectra classified C/2017 K2 as a typical-to-enriched comet, with HCN identified as the main parent molecule of CN, and C$_2$ probably originating from C$_2$H$_2$ rather than C$_2$H$_6$.}
  {}


\maketitle
\section{Introduction}
\label{sec_intro}

Comets are preserved relics from the primordial phases of the solar system's formation, having condensed from icy and dusty material surrounding the proto-Sun approximately 4.6 billion years ago. Consequently, they provide direct evidence of material from this era. Following their formation, significant gravitational interactions with the giant planets caused their dispersion into present reservoirs, namely the Kuiper Belt (the source of the ecliptic Jupiter Family Comets) and the Oort Cloud (the source of isotropic Long period comets). These frozen nuclei are believed to have retained the majority of their chemical and mineralogical properties associated with their region of origin within the protoplanetary disk, thereby offering invaluable insights into the initial phases of the solar system \citep{Gomes2005,Morbidelli2007}.

\begin{table*} 
		\begin{center}
			\caption{TRAPPIST observational circumstances and orbital elements of comet C/2017 K2. }
			\label{TRAPPIST-observation}
			\resizebox{\textwidth}{!}{%
				\begin{tabular}{lcccccccccc}
					\hline
					\hline
					Comet       &i          &T$_j$ &Peri  &Tp     &r$_h$-range &Dates     &\multicolumn{2}{c}{Number of nights}&\multicolumn{2}{c}{Number of images}\\
					              &($^\circ$) &      &(au)  &(UTC)  &Pre/Post (au)    &Start/End &TN&TS&TN&TS\\
					\hline
                     & & & & & & & & & &\\
	                C/2017 K2 (PanSTARRS)  &87.55 &0.17 &1.79 &19.12.2022&15.18 - 1.79 - 8.46 &25.10.2017 - 20.04.2025 &151 &120 &1323 &881\\
                 & & & & & & & & & &\\
					\hline	
					\hline
            \label{table1}
			\end{tabular}}
		\end{center}	
        {\footnotesize {\bf Note:} i: Inclination, T$_j$: Jupiter-Tisserand invariant, Peri: Perihelion distance, T$_p$: Time of perihelion passage. The format of the date is Day.Month.Year}	
	\end{table*}

The most abundant volatile in comets, water, can only sublimate at temperatures present within 3.0 au of the Sun, as originally suggested in the "dirty snowball" model of the nucleus \citep{Whipple1950}.
For comets active at greater distances, more volatile species such as CO or CO$_{2}$ must be sublimating, or another mechanism such as ice crystallization or non-thermal processes must be at play \citep{jewitt_2019K2}.

An example of such a distant active comet is C/2017 K2 (PanSTARRS), hereafter K2, an Oort cloud comet, discovered by the Pan-STARRS survey \citep{Kaiser2002} in May 2017, when it was at a heliocentric distance of r$_h$=16.1 au \citep{Wainscoat2017}. Prediscovery images of K2 were found showing the comet exhibiting activity at an extremely large distance of 23.8 au in May 2013 \citep{Jewitt2017,Meech2017,Hui2018}. Comet K2 is the second-most distant discovery of an active comet. At such a distance, the activity cannot be driven by the sublimation of water ice in the nucleus, which is the principal mechanism for comets in the inner solar system. The sublimation of super-volatile ice, including CO, CO$_2$, N$_2$ and O$_2$, drives the ejection of dust and gas, resulting in detectable cometary activity even at extreme distances from the Sun \citep{Jewitt2017}. Sub-millimeter observations later confirmed the presence of carbon monoxide (CO) in K2's coma with Q$_{CO}$ = (1.6 $\pm$ 0.5) e+27 at r$_h$ = 6.72 au \citep{Yang2021}.\\
\cite{Krolikowska2018}, performed dynamical simulations showing that K2 has probably never previously entered the inner solar system (r$_h$$\leq$5 au) where substantial sublimation on the nucleus can take place. Its original semi-major axis is now estimated to be as large as 28000 au \citep{Combi2025}, which, following the definition of \cite{A'Hearn1995} ($>$20000 au), should then be classified as a Dynamically New Comet (DNC). Photometric analyses have suggested that DNCs tend to behave differently from returning comets and often exhibit more asymmetric light curves that are substantially brighter before than after perihelion (\cite{Carrie2024PSJ}, \cite{HmiddouchEPSC24}). Such differences in photometric behavior may arise from intrinsic variations in the nucleus and coma properties affected by thermal processing in the inner solar system. K2's brightness while still in the outer Solar System makes it a particularly compelling target for evaluating the effects of solar heating on the properties of a relatively fresh comet \citep{Zhang2022}. It is also important to study the activity behavior of DNCs and composition from far away down to perihelion in the context of the new ESA F-class mission, Comet Interceptor \citep{Jones2024}. This mission will flyby a pristine comet visiting for the first time the inner solar system, to better understand those comets and their peculiarities.

\section{Observation and data reduction}
\label{sec:data_reduction}
\subsection{Photometry (TRAPPIST)}

We used both TRAPPIST\footnote{\url{https://www.trappist.uliege.be/}} (TRAnsiting Planets and Planetesimals Small Telescopes) North and South, hereafter TN and TS \citep{Jehin2011}, to observe and follow comet K2 for almost eight years. TRAPPIST-South is equipped with a 2K$\times$2K FLI Proline CCD with a pixel scale of 0.65 arcsec/pixel, resulting in a FOV of 22'$\times$22', while TRAPPIST-North is equipped with an Andor IKONL BEX2 DD (0.59 arcsec/pixel) with a 20'$\times$20' field of view.

The observations were carried out with a binning of 2X2, resulting in a plate scale of 1.30 and 1.20 arcsec/pixel, respectively. In addition to the standard Johnson-Cousin B, V, Rc, and Ic broad-band filters, the telescopes are also equipped with cometary HB narrow-band filters that were specifically designed for observing comet Hale-Bopp \citep{Farnham2000}. These narrowband filters isolate emissions from OH (309.7 nm), NH (336.1 nm), CN (386.9 nm), C$_3$ (406.3 nm) and C$_2$ (513.5 nm), as well as three emission-free wavelength regions (BC at 445.3 nm, GC at 525.9 nm and RC at 713.3 nm). 

Our observations of comet K2 began with the TN telescope on October 25, 2017, using broad-band filters. At that time, the comet was at a distance of 15.18 au from the Sun and had a visual magnitude of 19.7. We continued observing the comet with both broadband and HB narrowband filters using the TS telescope starting on September 9, 2021, when it became visible and bright enough from the southern hemisphere (r$_h$=5.4 au). We monitored the comet about two times a week until October 24, 2022, when it became too low in the sky and was not visible due to its solar conjunction. After perihelion on December 19, 2022 (r$_h$ = 1.79 au), we recovered the comet on January 27, 2023. We continued observing it as long as it remained bright and high enough in the sky, until May 3, 2023 (r$_h$=2.46 au). Then we continued to monitor its activity until April 2025 (r$_h$=8.46 au) with only the broad-band filters.

Overall, we collected about 2204 broad-band images and 174 narrow-band images of the comet over a total of 271 nights, spending 8 years (Table \ref{TRAPPIST-observation}). The exposure times ranged from 30 to 240 seconds for the broad-band filters and from 300 to 1500 seconds for the narrow-band filters based on the brightness of the comet as it approached or moved away from the Sun. The data were calibrated using standard procedures, such as subtracting bias, dark, and performing flat-field correction. The sky contamination was then removed, and the data were flux calibrated using standard stars observed regularly during the same period, using the same procedure explained in previous TRAPPIST publications \citep{Opitom2015a, Moulane2018}. After the determination of the comet's photocenter, we derived median radial brightness profiles from the gas and dust images. Then, we removed dust contamination from the gas radial profiles using images of comets in the BC filter, because it is less contaminated by cometary gas emission than the other dust filters \citep{Farnham2000}. The fluxes of OH, NH, CN, C$_3$, and C$_{2}$ were converted into column densities, and we fitted a Haser model \citep{Haser1957} to the profiles to derive the production rates. The Haser model, although not physically accurate as it assumes the one-step photo-dissociation of the parent molecule into daughter molecules in a spherically symmetric coma, is commonly used to determine the gas production rates from optical comet observations. It enables comparisons between observations made by various observers, as well as comparisons between different comets. The model adjustment is performed at a nucleocentric distance of around 10,000 km to avoid PSF and seeing effects around the optocenter. At greater nucleocentric distances, the signal usually becomes fainter, especially in the OH filter, for which the signal-to-noise ratio (S/N) is lower. Fluorescence efficiencies (also called g-factors) from David Schleicher’s website\footnote{\url{https://asteroid.lowell.edu/comet/gfactor.html}} (see also Table \ref{g-factors_table}) were used to convert fluxes into column densities. The C$_2$ g-factor is determined only by considering C$_2$ in a triplet state, as noted by \cite{AHearn1982}. Similarly, the C$_3$ g-factor were also determined by \cite{AHearn1982}. CN and NH fluorescence efficiencies vary with both the heliocentric distance and the velocity and are taken, respectively, from \cite{Schleicher2010} and \cite{Meier1998}. The value of the g-factor for the OH (0-0) band centered near 3090 \text{\AA} varies with both the heliocentric distance and velocity \citep{SchleicherandAHearn1988}. We used scale lengths from \cite{A'Hearn1995} scaled as r$_h$$^2$. The choice of using scale lengths, with r as heliocentric distance, was made to facilitate comparison with various data sets, particularly the extensive data set from \cite{A'Hearn1995}.

We used broad-band filters including B, V, Rc, and Ic \citep{Bessell1990} to monitor the evolution of the light curve in various colors over almost eight years. We used observations with the narrow-band BC, GC, and RC filters and also with the broad-band Rc filter to estimate dust production. From the dust profiles, we derived the Af$\rho$ parameter, as first introduced by \cite{A'Hearn1984}. All Af$\rho$ values were corrected for the phase angle to obtain A(0)f$\rho$. Several phase functions have been proposed (\citealp{Divine1981}; \citealp{Hanner1989}; \citealp{Schleicher1998} or \citealp{Marcus2007}). We used the phase function described by Schleicher \footnote{\url{https://asteroid.lowell.edu/comet/dustphase.html}}, which is a combination of two different phase functions of \citealp{Schleicher1998} and \citealp{Marcus2007}.

\subsection{Optical spectroscopy with UVES at UT2/VLT}
To conduct a detailed investigation of the composition of K2 in the optical range and its evolution while getting closer to perihelion, we conducted a program with the high-resolution Ultraviolet-Visual Echelle Spectrograph (UVES) on the Unit 2 telescope (UT2) of the ESO 8-m VLT, at Paranal on three different epochs before and after the water sublimation line ($\sim$ 3 au). We used two distinct UVES standard settings to achieve a complete optical range coverage (303-1060 nm): the dichroic \#1 (346 + 580), which goes from 303 to 388 nm in the blue and 476 nm to 684 nm in the red, and the dichroic  \#2 setting (437 + 860), covering 373 to 499 nm in the blue and 660 to 1060 nm in the red. The details of these spectral observations are summarized in Table \ref{tab:crires_and_uves_obs_log}. We used a 0.45" $\times$ 10" slit in the blue, resulting in a resolution power of $\sim$ 70,000, and a 0.45" $\times$ 12" slit in the red, resulting in a resolution power of $\sim$ 90,000. The slit was centered on the inner coma and aligned with the Sun direction.

We processed the data using the ESO UVES pipeline \citep{Ballester2000}, supplemented with custom routines for extraction and cosmic ray removal, followed by correction for the Doppler shift due to the comet's relative velocity to Earth. The spectra were calibrated in absolute flux using either the archived master response curve or a response curve derived from a standard star observed near the science spectrum. Finally, the continuum, including sunlight reflected by cometary dust grains, was removed using the BASS2000\footnote{\url{https://bass2000.obspm.fr/solar\_spect.php}} solar spectrum, adjusted to match the comet slope. Consequently, the final spectrum includes only the gas component. More details on the data reduction process are available in \cite{Manfroid2021Natur} and references therein.

\subsection{NIR spectroscopy with CRIRES$^+$ at UT3/VLT}
CRIRES$^+$, mounted on the Unit 3 telescope (UT3) at the VLT, is the ESO high-resolution infrared \( 0.95\text{--}5.3 \, \mu\text{m} \) spectrograph. This upgraded version of the original CRIRES instrument \citep{Kaeufl2004, Dorn2014Msngr}, is now a cross-dispersed echelle spectrometer, offering ten times greater wavelength coverage than its previous version while maintaining a high spectral resolving power of 40,000 for a slit width of 0.4''. The enhanced CRIRES$^+$ is equipped with three new detectors, providing a larger field coverage area, lower noise, higher quantum efficiency, and reduced dark current. Performance is further enhanced by the multi-application curvature adaptive optic system (MACAO) \citep{Paufique2004}. 

\begin{table}[h]
\begin{center}
\caption{The simultaneous UVES and CRIRES$^+$ observational circumstances of comet C/2017 K2, with the UT2 and UT3 telescopes of the ESO VLT.}
\label{obs_UVES_CRIRES_and_TRAPPIST}
\renewcommand{\arraystretch}{1.1}
    \normalsize
\resizebox{0.35\textwidth}{!}{%
\begin{tabular}{llccc}
\hline	
\hline
Date 2022 & r$_h$(au)    &r$_{\Delta}$(au)  &$\dot{r_{\Delta}}$ (km/s) \\ 
\hline 
May 9  &  3.23 &  2.66 & -37.7 \\
July 5  &  2.73 &  1.82 & -6.65 \\
Sept 21  &  2.12 &  2.32 & 15.98  \\
Sept 22  &  2.11 &  2.35 & 15.6 \\
\hline
\hline
\end{tabular}}
\end{center}

\label{table2}
\end{table}

Observations of K2 were conducted using CRIRES$^+$ over three nights, simultaneously with UVES observations, as detailed in Tables \ref{obs_UVES_CRIRES_and_TRAPPIST}, and \ref{tab:crires_and_uves_obs_log}. The settings were selected to capture the majority of primary volatiles (e.g., H\(_2\)O, CO, C\(_2\)H\(_6\), CH\(_4\), HCN, NH\(_3\), etc.) and to monitor their evolution as the comet approaches the Sun \citep{Lippi2023}. We used a slit of 0.4$\arcsec$, aligned along the extended Sun-comet radius vector.

Data were processed using custom semi-automated procedures \citep[see][]{Villa2011, VillaPSG2022, 2005PhDT.......270B}, enabling efficient spectral analysis. Spectral calibration and correction for telluric absorption were performed by comparing the data to highly accurate atmospheric radiance and transmittance models generated with PUMAS/PSG \citep{Villanueva2018}. Flux calibration was achieved using the spectra of a standard star observed close in time to the comet and processed with the same algorithms. Production rates and relative abundances (i.e., mixing ratios relative to water) of various primary species in the coma were determined as described in \cite{Lippi2020} (and references therein), using advanced fluorescence models (e.g., \cite{Villanueva2012}, \cite{2011ApJ...729..135R}). Figure \ref{fig:CRIRES_spectra} shows a selection of K2 spectra acquired with CRIRES$^+$.  

\begin{table}[h!] 
  \caption{CRIRES$^+$ and UVES observation log}
  \label{tab:crires_and_uves_obs_log}
  \centering
  \resizebox{1\columnwidth}{!}{%
  \begin{tabular}{ccccc}
    \hline
    \hline
    Date & Instrument & Setting & Airmass & Exposure time [s] \\
    \hline
        9 May. 2022
        & CRIRES$^+$ & L3377 & 1.130 & 2280 \\
        & CRIRES$^+$ & M4318 & 1.115 & 1400\\
        & UVES       &  346+580 & 1.24 & 3000 \\
        & UVES       & 437+860 & 1.28 & 3000 \\
    \hline
    5 July 2022 
        & CRIRES$^+$ & L3377 & 1.130 & 3600 \\
        & CRIRES$^+$ & M4318 & 1.115 & 2880 \\
        & UVES (a)      & 437+860 & 1.28 & 3000 \\
        & UVES (a)     & 346+580 & 1.15 & 3000 \\
        & UVES (b)     & 346+580  & 1.11 & 3600 \\
        & UVES (b)       & 437+860 & 1.13 & 3600 \\
    \hline
    21 Sep. 2022 
        & CRIRES$^+$ & L3377 & 1.375 & 2400 \\
        & CRIRES$^+$ & M4318 & 1.720 & 1600 \\
        & UVES       & 346+580 & 1.34 & 3000 \\
        & UVES       & 437+860 & 1.71 & 3000 \\
    \hline
    22 Sep. 2022 
        & CRIRES$^+$ & L3377 & 1.595 & 2640 \\
        & CRIRES$^+$ & L3302 & 1.288 & 2880 \\
        & UVES       & 346+580  & 1.29 & 3600 \\
        & UVES       & 437+860 & 1.66 & 3600 \\
    \hline
    \hline
    \\[2pt] 
    \multicolumn{5}{l}{\footnotesize {\bf Note:}(a) and (b) are two different UVES observations executed during the same night.}\\
  \end{tabular}%
}
\end{table}

\section{Data analysis and results}
\label{sec_analysis}
\subsection{Photometry (TRAPPIST)}

In this section, we present the comet K2 light curves in different filters as well as the evolution of its activity and composition before and after perihelion.

\subsubsection{Light curves and coma dust colors}
Figure \ref{fig:mag_K2}, presents the evolution of the magnitude in a 5 arcsec aperture, using the B, V, Rc, and Ic broadband filters compared to the magnitude reported in the JPL ephemeris. The data are fitted using the standard magnitude formula "$M = M_{0} + 5 \times \log_{10}(r_{\Delta}) + 2.5 \times n \times \log_{10}(r_{h})$", using the best fit parameters; n is a coefficient and $M_{0}$ the absolute magnitude. $r_{\Delta}$ and $r_{h}$ are the geocentric and heliocentric distances, respectively. K2 was observed in the four-band filters for its entire inner solar system passage. The comet reached its peak brightness, corresponding to an Rc-band magnitude of 11.18 on January 30, 2023. The light curve shows large-scale deviations but no outbursts. It was not possible to observe the comet the month before and after perihelion due to its conjunction with the Sun.

\begin{figure}[h!]
	\centering	\includegraphics[width=0.98\linewidth]{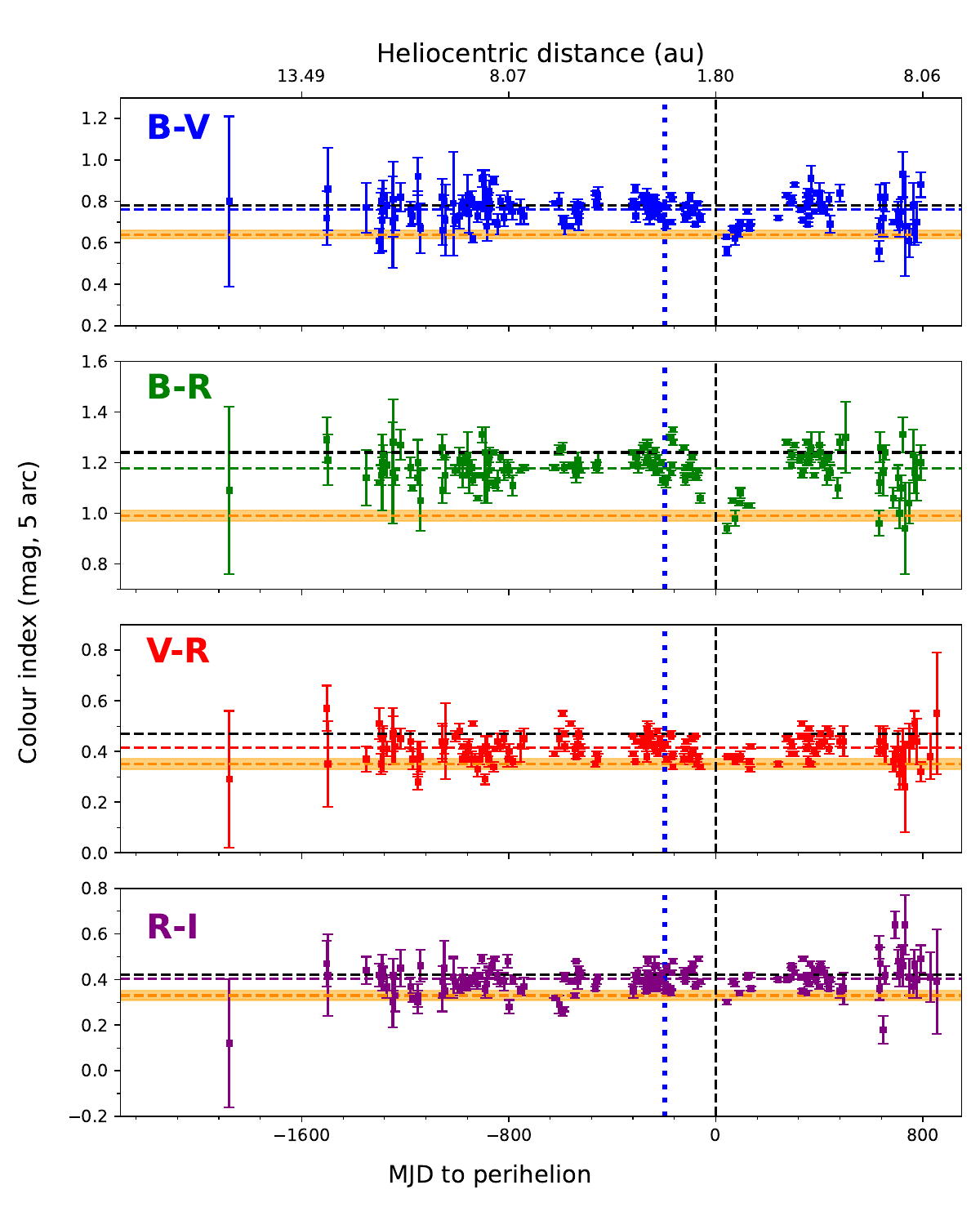}
	\caption{B-V, B-R, V-R, and R-I colors of C/2017 K2 vs time to perihelion and heliocentric distance, with the mean values from Table \ref{colors_table} compared with the average value of 25 active LPC from \cite{Jewitt2015} (black horizontal dashed lines), and the Sun colors from \cite{Holmberg2006} (orange horizontal dashed lines). The vertical dashed line represents the perihelion, and the vertical blue dotted line represents the water ice sublimation boundary ($\sim$ 3 au).} 
	\label{fig:colors_K2}
\end{figure}

\begin{figure}[h]
	\centering
	\includegraphics[width=0.9\linewidth]{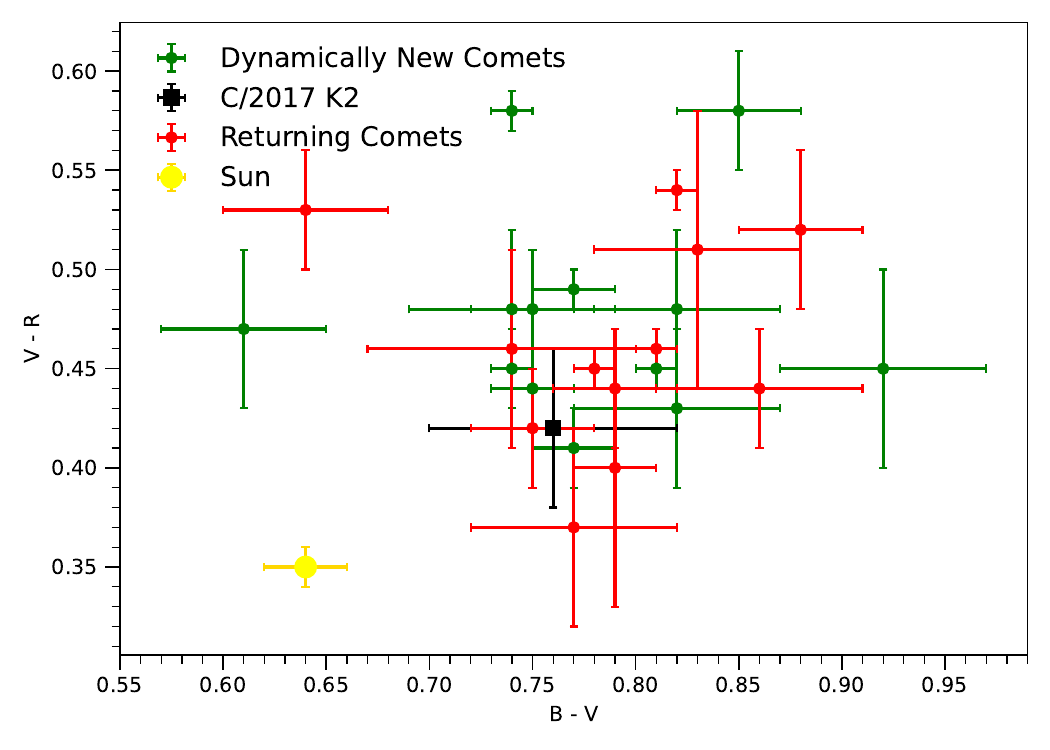}
	\caption{Color–color plot comparing C/2017 K2 (square) with dynamically new comets (green) and returning comets (red) from \cite{Jewitt2015}. The color of the Sun is marked by a yellow circle.}
	\label{fig:allcolors_K2}
\end{figure}

Figure \ref{fig:colors_K2} shows the color indices B-V, B-R, V-R, and R-I, or colors in short, of K2 as a function of the heliocentric distance. The colors are surprisingly constant throughout the whole range of heliocentric distances, except near the perihelion ($<$ 2.0 au) where the B and V filters are contaminated by the gaseous emission of CN and C$_2$ respectively. This means that the properties of the dust (such as grain size) do not change much in the small aperture centered on the nucleus. The colors agree very well with those measured for 25 long-period active comets (LPC) by \cite{Jewitt2015} (see Table \ref{colors_table}, and Figure \ref{fig:allcolors_K2}). We divided the comets in \cite{Jewitt2015} into two groups: 13 dynamically new comets (DNCs) and 12 returning comets (RCs), and compared their broadband colors with those of comet K2 (see Figure \ref{fig:allcolors_K2}). Our analysis reveals no significant color differences between the two dynamical classes, which is also consistent with the result from \cite{Carrie2024PSJ} for 21 LPCs. The colors do not allow for distinguishing comets from different dynamical origins. We may note that K2 lies on the blue side of the distribution of colors.

\begin{table}[h!]
\begin{center}
{\renewcommand{\arraystretch}{3}
\caption{Comparison of C/2017 K2 colors with active LPCs}
\label{colors_table}
    \renewcommand{\arraystretch}{2}
    \normalsize
\resizebox{0.48\textwidth}{!}{%
\begin{tabular}{llccccc}
\hline	
\hline
Object & B - V    & V -R & R - I & B - R\\ 
\hline 
C/2017 K2 & 0.76$\pm$0.06 & 0.42$\pm$0.04 & 0.40$\pm$0.05  & 1.18$\pm$0.06\\
DNCs & 0.76$\pm$0.05 & 0.49$\pm$0.05 & 0.42$\pm$0.02 & 1.26$\pm$0.07\\
RCs & 0.80$\pm$0.05 & 0.48$\pm$0.05 & 0.37$\pm$0.09 & $ 1.28\pm$0.05 \\
Active LPCs & 0.78$\pm$0.02 & 0.47$\pm$0.02 &0.42$\pm$0.03 &1.24$\pm$0.02 \\

\hline
Sun  & 0.64$\pm$0.02 & 0.35$\pm$0.01 & 0.33$\pm$0.01 & 0.99$\pm$0.02\\
\hline
\hline
\end{tabular}}}
\end{center}
        {\footnotesize {\bf Note:} The mean colors of C/2017 K2 are compared to the mean colors of DNC and returning comets from the sample of 25 LPC from \cite{Jewitt2015}. The Sun color is from \cite{Holmberg2006}.}	
\end{table}

\begin{figure}
	\centering	\includegraphics[width=0.9\linewidth]{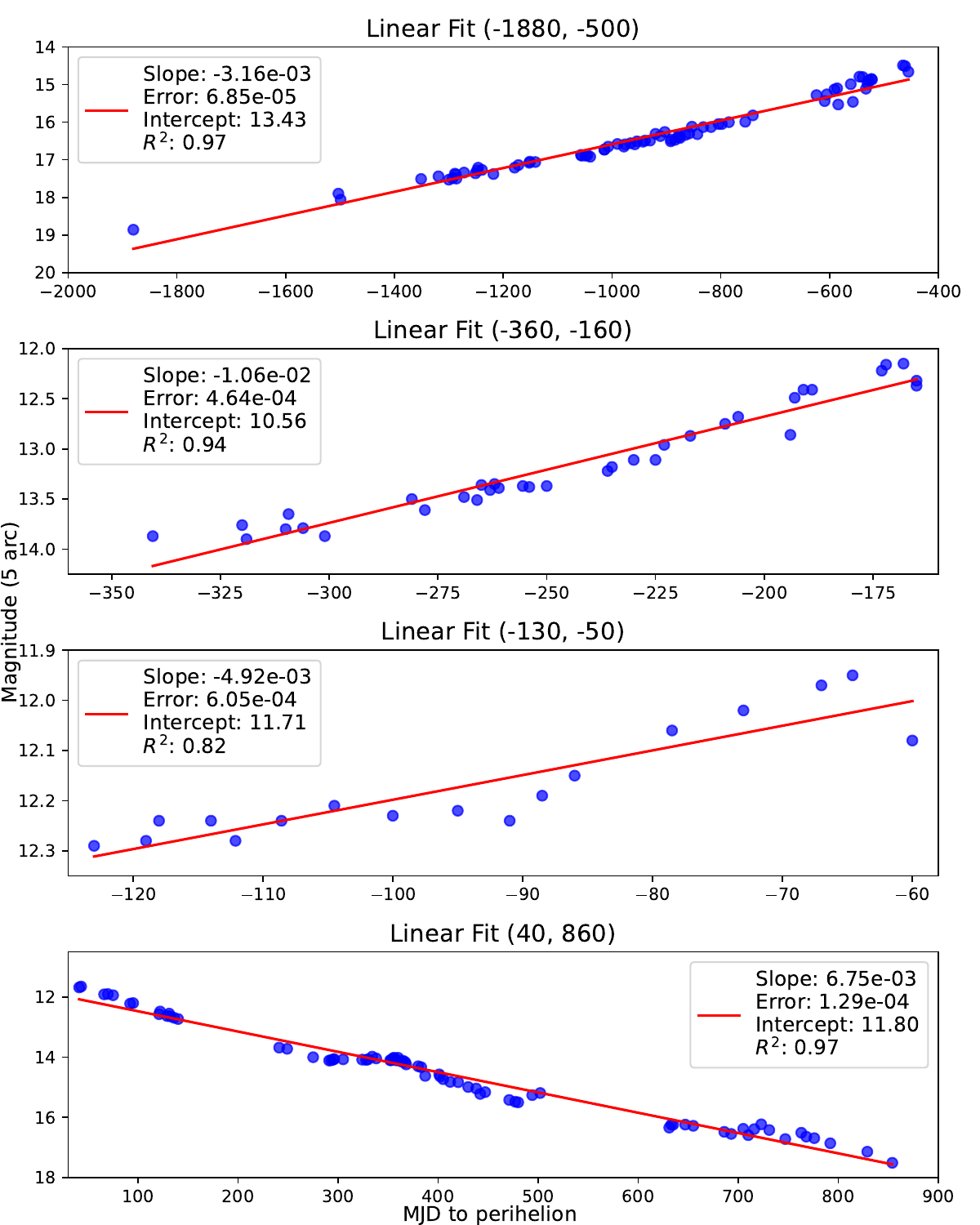}
	\caption{The four different slope regimes in the Rc light curve of comet C/2017 K2.} 
	\label{slopes_plot}
\end{figure}

\begin{table}[h!]
    \centering
    \caption{Slope fitting results in the light curve evolution of comet C/2017 K2.}
    \label{slopes_tab}
    \setlength{\tabcolsep}{2pt}
    \renewcommand{\arraystretch}{1.46}
    \normalsize
    \resizebox{\columnwidth}{!}{
    \begin{tabular}{lccccc}
        \hline
        \hline
        Days to perihelion &r$_h$-range (au) & Best Slope & Intercept & R-squared \\
        \hline
        -1880, -500 & 15.18 - 5.3 & -3.16 $\pm$ $0.068 \times 10^{-3}$ & 13.43 & 0.97 \\
        -360, -160 & 4.29 - 2.71  & -1.06 $\pm$ $0.046 \times 10^{-2}$ & 10.56 & 0.94 \\
        -130, -50 & 2.37 - 1.96  & -4.92  $\pm$ $0.60 \times 10^{-3}$ & 11.71 & 0.82 \\
        40, 860 & 1.87 - 8.46  & 6.75  $\pm$ $0.13 \times 10^{-3}$  & 11.80 & 0.97 \\
        \hline
        \hline
\end{tabular}}
\end{table}
\begin{figure*}[h!]
	\centering	\includegraphics[width=0.8\linewidth]{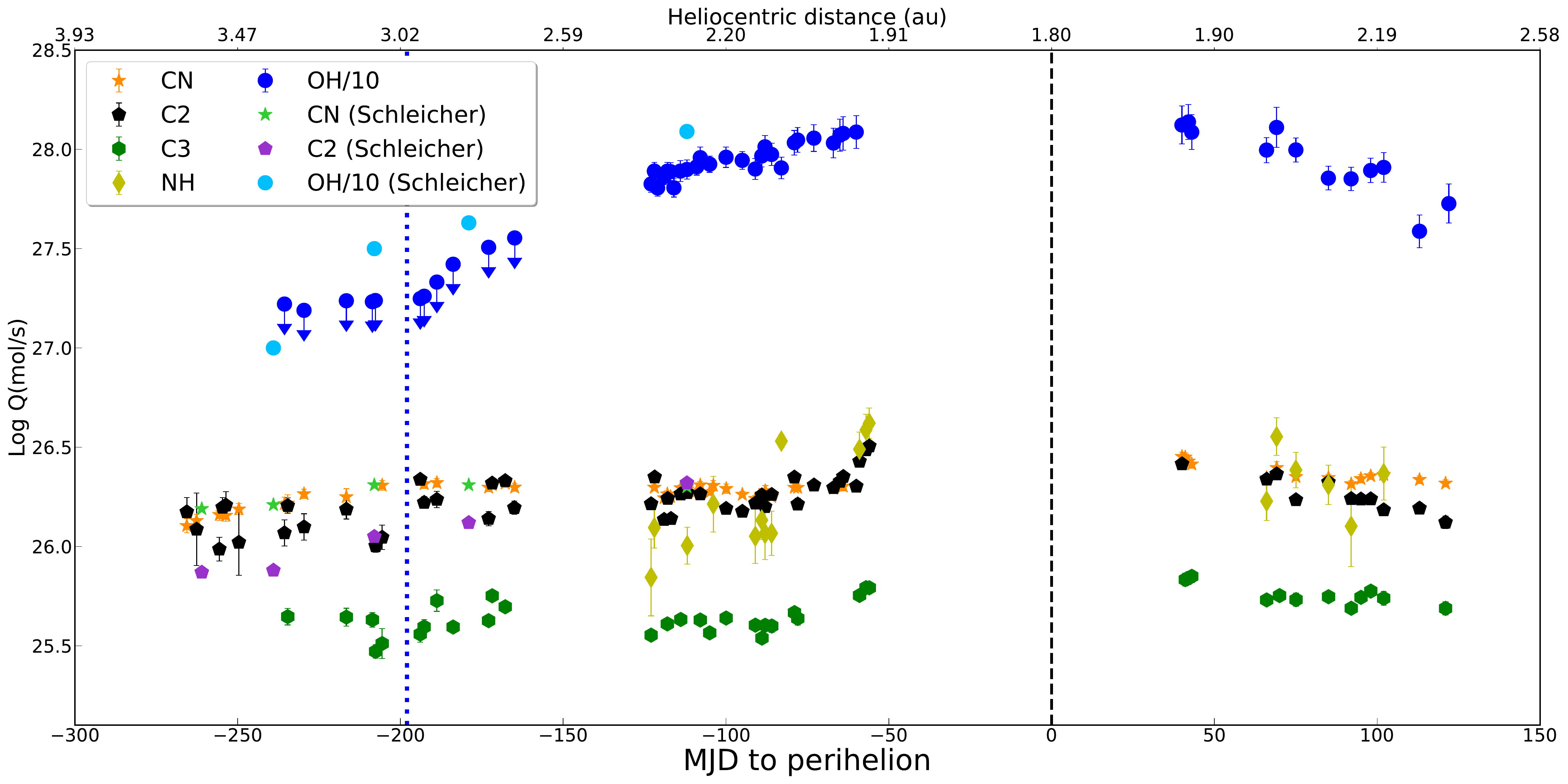}
	\caption{OH, NH, CN, C$_2$, and C$_3$ logarithmic production rates of comet C/2017 K2 from TRAPPIST photometry as a function of time and the heliocentric distance. The vertical dashed line indicates the perihelion at 1.79 au on December 19, 2022.} 
	\label{fig:prK2}
\end{figure*}
To investigate the temporal evolution of the comet's photometric activity, we performed linear fits over four distinct time intervals relative to perihelion (see Figure \ref{slopes_plot}, and Table \ref{slopes_tab}). Each interval corresponds to a different phase in the comet's activity:\\
Distant pre-perihelion phase from –1880 to –500 days (from 15.18 to 5.3 au): The comet shows a slow and steady brightening trend. This behavior likely corresponds to the early onset of activity, dominated by the sublimation of highly volatile ices (e.g, CO, CO$_2$) at large heliocentric distances. The activity is weak, but measurable, with a low negative slope indicating increasing brightness.\\
Approaching perihelion from –360 to –160 days (from 4.29 to 2.71 au): A sudden and more pronounced brightening is observed, reflected by a steeper negative slope. This phase is characterized by a rapid increase in activity, likely driven by the onset of water-ice sublimation and increased dust production. The higher correlation (R²) suggests that the linear model fits this interval well, indicating a relatively smooth ramp-up in activity.\\
Near-perihelion plateau from –130 to –50 days (from 2.37 to 1.96 au): The fit yields a nearly flat slope, indicating a stable and plateau phase in the brightness and activity of the comet. This disappointing performance of the comet, which was expected to be brighter at perihelion, was noticed by many. This may suggest a temporary equilibrium between solar input and gas/dust production. The low correlation indicates the presence of more complex or possibly non-linear behavior during this transitional phase.\\
Post-perihelion decline from +40 to +460 days (from 1.87 to 8.46 au): After perihelion, the comet shows a larger slope and a fast fading. This decline is expected as the heliocentric distance increases and solar heating diminishes. The linear fit captures this trend well, with a consistent decrease in brightness over time. We note only one slope after perihelion.

\subsubsection{Gas production rates, Af$\rho$ and their ratios}
Figure \ref{fig:prK2} illustrates the production rates of several volatile species (OH, NH, C$_3$, CN, and C$_2$) in comet K2. The x-axis denotes the time to perihelion and the heliocentric distance, with a vertical dashed line indicating the perihelion at 1.79 au. The y-axis displays the logarithmic production rates in molecules per second, representing the outgassing rates for each species. For visibility, the OH production is scaled down by a factor of 10. The logarithmic scale facilitates the comparison of species with different production levels and highlights the comet’s changing activity as it approaches or moves away from the Sun. The production rate values for the various species are summarized in Table \ref{tab:K2Qs}. We first detected CN and C$_2$ radicals at the end of March 2022 at 3.62 au, followed by the majority of other radicals that appeared in the coma about a month later, except for NH, which was observed only in mid-August 2022. The production rates gradually increased as the comet approached the Sun, from 3.62 to 2.71 au, and then it surprisingly stabilized, showing a plateau-like behavior both before and after perihelion, similar to the optical light curve.

\begin{figure}[h]
	\centering
	\includegraphics[width=0.98\linewidth]{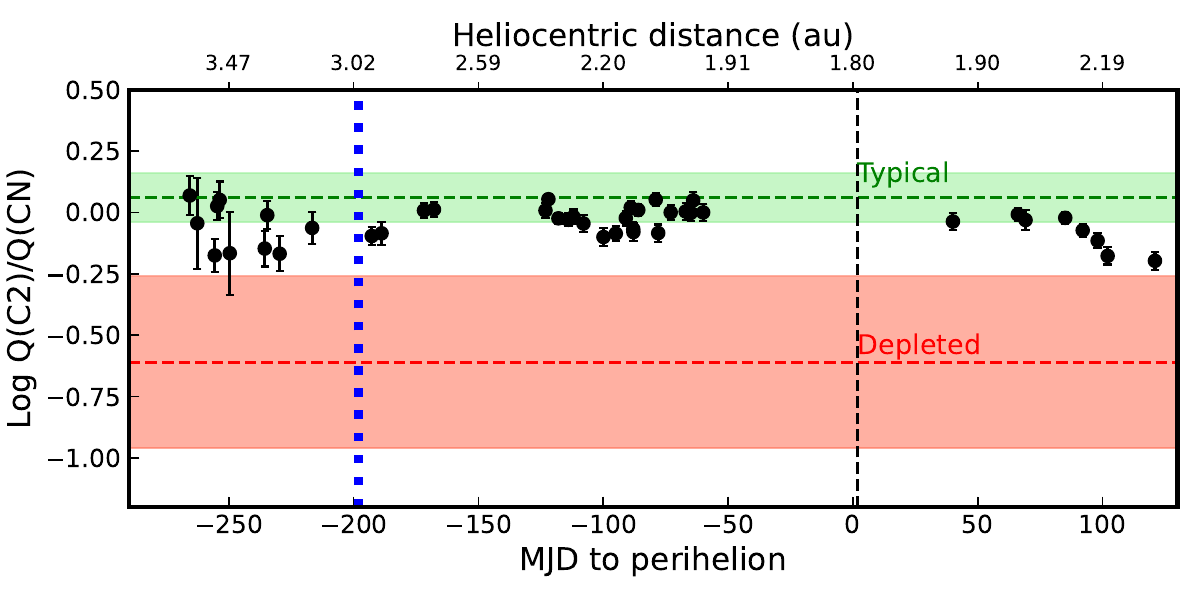}
	\caption{The logarithm of  C$_2$/CN production rate ratios of comet C/2017 K2, as a function of time and the heliocentric distance.} 
	\label{C2_CN_ratios}
\end{figure}

\begin{table}[h!]
	\begin{center}
		\caption{Average logarithmic production rates and A(0)f$\rho$ ratios of Comet C/2017 K2 in comparison with \cite{A'Hearn1995} Taxonomic Classes.}
			\label{tab:K2:abundance}
            \renewcommand{\arraystretch}{1.5} 
            \normalsize
			\resizebox{0.45\textwidth}{!}{%
			\begin{tabular}{lccc}
				\hline	
				\hline
				& C/2017 K2 & Typical comets & Depleted comets \\
				& \multicolumn{1}{c}{ (This work)} & \multicolumn{2}{c}{\citep{A'Hearn1995}} \\ 
				\hline   
				C$_2$/CN             &-0.037$\pm$0.005        &  0.06$\pm$0.10 &-0.61$\pm$0.35  \\
                C$_2$/OH             &-2.65$\pm$0.07        & -2.44$\pm$0.20 &-3.30$\pm$0.35  \\
				CN/OH                &-2.58$\pm$0.34        & -2.50$\pm$0.18 &-2.69$\pm$0.14  \\
				C$_3$/OH             &-3.29$\pm$0.07        & -3.59$\pm$0.29 &-4.18$\pm$0.28  \\
				NH/OH                &-2.83$\pm$0.14        & -2.37$\pm$0.27 &-2.48$\pm$0.34  \\
				A(0)f$\rho$/OH       &-24.75$\pm$0.43        & -25.82$\pm$0.40 &-25.30$\pm$0.29 \\
				\hline	
				\hline
	\end{tabular}}
	\end{center}
 {\footnotesize {\bf Note:} The A(0)f$\rho$ values used were computed from the images taken with the RC narrow-band filter.}	
\end{table}

\begin{figure*}[h!]
	\centering	\includegraphics[width=0.8\linewidth]{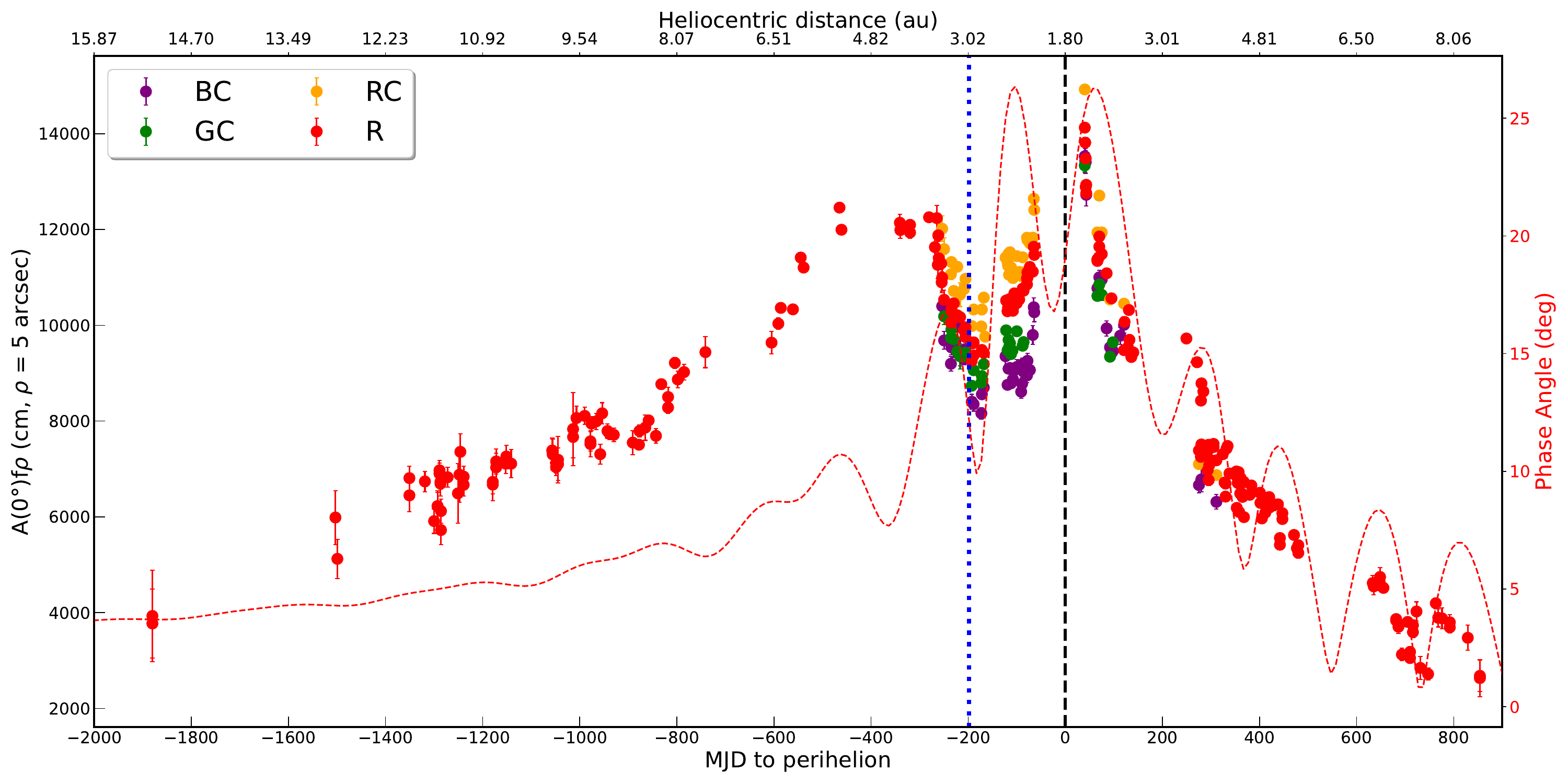}
	\caption{The A(0)f$\rho$ parameter of comet C/2017 K2 from the broad- and narrow-band filters as a function of the heliocentric distance.} 
	\label{fig:afrhoK2}
\end{figure*}

\begin{figure}[h]
	\centering
	\includegraphics[width=0.95\linewidth]{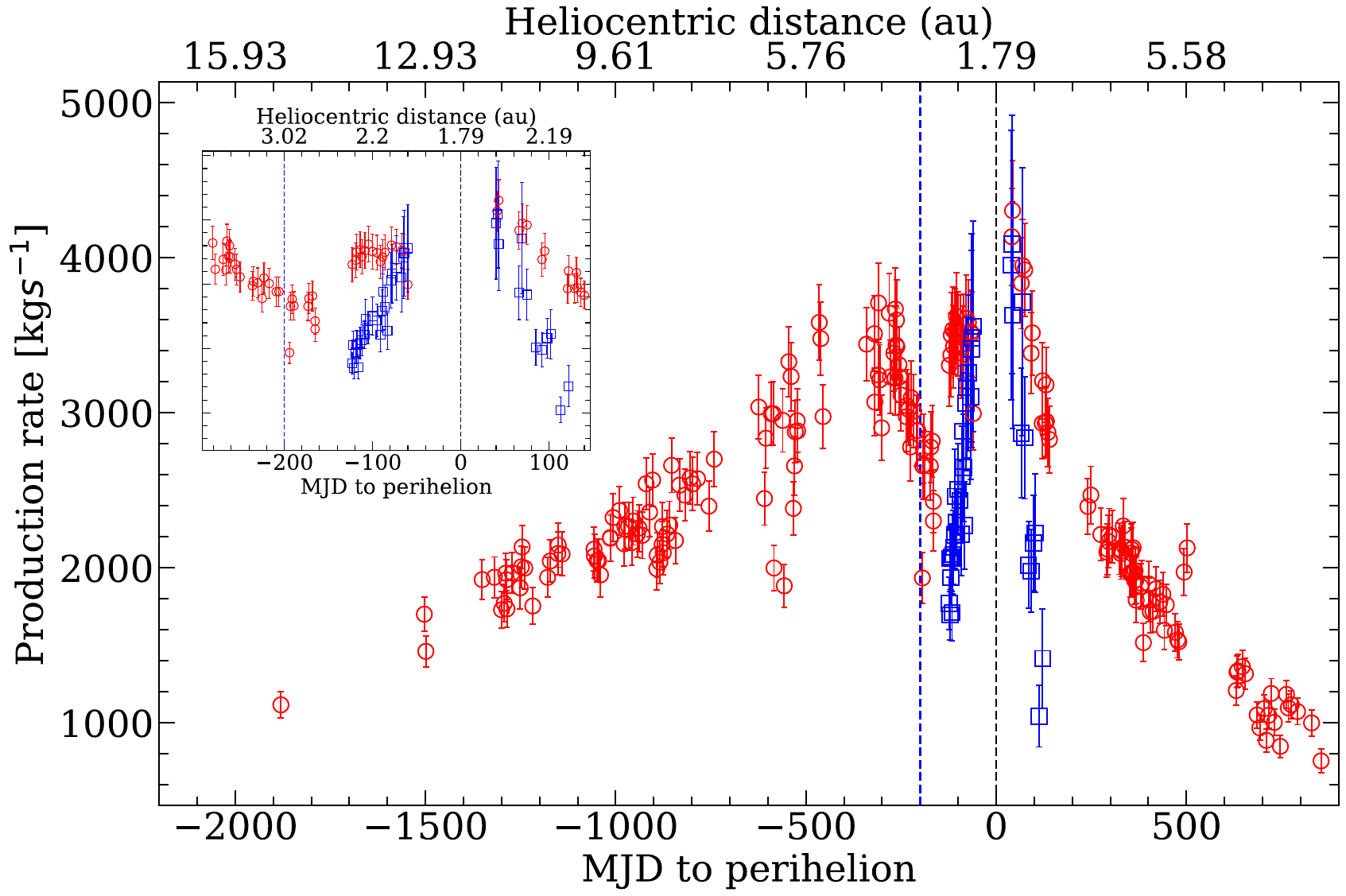}
	\caption{The comparison of dust mass loss and water mass production in comet K2 as a function of days to perihelion. The dust mass loss was computed using equation \ref{equation3}, and water mass production was computed from the OH production rate observed by TRAPPIST. The vertical blue dotted line represents the water ice sublimation boundary ($\sim$ 3 au) \citep{womack2017, Crovisier2000}, within which the main driving source of comet outgassing changes from supervolatile ices to H$_2$O ice \citep{kwon_k2}. The vertical dashed line represents the perihelion.} 
	\label{17K2_dust_water_prod}
\end{figure}

CN remained detectable in our data until mid-October 2023 at 3.88 au, while C$_2$ and C$_3$ were no longer observed after early October at 3.83 au. OH was detected until September 20 at 3.69 au, while NH was detected until the end of March 2023 at 2.20 au. The OH, CN, and C$_2$ values are in good agreement with \cite{Combi2025} for the water production rates from SOHO measurements of the Lyman-alpha H line before and after perihelion, and with Schleicher's narrow band photometric measurements (private communication, see also Figure \ref{fig:prK2}).

We calculated production rate ratios relative to CN and OH, as well as the dust-to-gas ratio. Comet K2 has a log [A(0)f$\rho$/Q(OH)] = -24.75 $\pm$ 0.43, which places K2 in the dust-rich regime, compared to typical cometary values (Table \ref{tab:K2:abundance}) reported by \cite{A'Hearn1995}. Figure \ref{C2_CN_ratios} illustrates the evolution of the logarithm of C$_2$/CN over time to the perihelion and heliocentric distances. According to the taxonomic classification by \cite{A'Hearn1995}, the comet falls into the 'typical' group, which is defined by a characteristic abundance of C$_2$ and C$_3$ relative to CN and OH. Table \ref{tab:K2:abundance} summarizes the production rates ratios in K2, compared to the comet database given in \cite{A'Hearn1995}.\\ 

The apparent magnitude of a comet depends on its heliocentric distance, geocentric distance, and phase angle at the time of observation. To remove the influence of geometry and illumination effects, we computed the absolute magnitude (H), defined as the magnitude that the comet would have if observed at a heliocentric and geocentric distance of 1 au and a phase angle of 0$^\circ$. It is given by:
\begin{equation}\label{abs_mag}
H = m - 5\log_{10}(r_h \times r_{\Delta}) - f(\alpha),
\end{equation}
where $m$ is the apparent magnitude in a given filter, $r_h$ is the heliocentric distance, $r_{\Delta}$ is the geocentric distance. The phase correction $f(\alpha)$ is defined as -2.5log$_{10}$[$\phi(\alpha)$], where $\phi(\alpha)$ is the phase function corresponding to the phase angle at the time of observation, as defined in \cite{Schleicher1998}. Figure \ref{fig:absmag_K2} illustrates the variation in absolute magnitude of comet K2 as a function of days and distance to the perihelion.

The absolute magnitude in the Rc band can be used to derive the effective scattering cross-section ($C_e$) \citep{jewitt_2019K2}, which provides insights into the dust content and activity of the comet. It is calculated as:
\begin{equation}
C_e = \frac{\pi r_0^2}{p} \cdot 10^{0.4[m_{\odot,R} - H_R]},
\label{equation2}
\end{equation}
where $r_0$ is the mean Earth–Sun distance in km, $p$ is the geometric albedo of the dust, and $m_{\odot,R}$ is the apparent magnitude of the Sun in the Rc band. Using $r_0 = 1.5 \times 10^8$ km and $m_{\odot,R} = -26.97$ \citep{sun_mag}, this simplifies equation \ref{equation2} to $C_e = (1.5\times10^6/p)\times10^{-(0.4H_R)}$.
For this work, we adopt an albedo value of $p = 0.04$ based on estimates from \cite{jewitt_2019K2} and \cite{zhang_2019K2}.

The effective scattering cross-section is further used to estimate the average dust mass loss rate via:
\begin{equation}
\frac{dM}{dt} = \frac{4}{3} \cdot \frac{\rho \bar{a} C_e}{\tau_r},
\label{equation3}
\end{equation}
where $\rho$ is the bulk density of the dust particles, $\bar{a}$ is the mean particle radius, and $\tau_r = L/v_{ej}$ is the residence time of the particles within an aperture of radius $L$, with $v_{ej}$ being the ejection velocity. We adopt $\rho = 500$ kg/m$^3$ and $\bar{a} = 100~\mu$m, following \cite{jewitt_2019K2}. A projected aperture radius of 5\arcsec was used for the analysis, corresponding to a physical diameter that varies with the comet’s geocentric distance. Although dust velocity varies with heliocentric distance, for comparative purposes, we use a fixed ejection velocity of 14 m/s consistent with the average velocity for 100~$\mu$m grains reported by \cite{Bin_Liu_22E3}.

On carefully analysing the variation of absolute magnitude as shown in Figure  \ref{fig:absmag_K2}, K2 displayed an unusual activity pattern as it approached the inner solar system. Initially, the comet appeared exceptionally bright at large distances ($\sim$15 au), exhibiting a low absolute magnitude, which suggests strong activity even in the outer solar system. This brightness was primarily driven by CO sublimation \citep{Meech2017}, which is consistent with the behaviour of the comet between 15 and 13 au, where the absolute magnitude decreased (the comet brightened). The same trend has also been reported by \cite{jewitt_2019K2} for observations at a similar heliocentric range. However, below 12 au, the absolute magnitude began to increase (the comet became fainter), indicating a decline in CO-driven activity. The most striking feature is the sharp increase in magnitude (fading) between 4 au and 3 au (see inset in Figure.\ref{fig:absmag_K2}). This can be attributed to the depletion of near-surface CO and/or CO$_2$, which was the main driver of activity until then. Moreover, the non-detection of OH emission in the TRAPPIST and UVES observations before the comet crossed 3 au implies that water (H$_2$O) sublimation had not yet begun, creating a temporary lull in activity. Furthermore, in May (r$_h$=3.23 au), we did not see strong CO and H$_2$O lines in CRIRES$^+$ spectra, which is in line with this hypothesis.

Once the comet approached closer to 3 au, the sublimation zone of water ice, the magnitude started to decrease again (brightening), signalling the onset of H$_2$O-driven activity. However, the comet did not become as bright as expected, as water production alone could not compensate for the loss of CO- and CO$_2$- driven activity. The activity trend in A(0)f$\rho$ and gas production provides further evidence for this interpretation. Figure \ref{fig:afrhoK2} shows the evolution of the dust proxy parameter Af$\rho$ in cm \citep{A'Hearn1984} (see Table \ref{tab:K2Qs}), measured using the broad-band dust continuum filter (Rc), and the narrow-band dust continuum filters (RC, GC, BC). The vertical dashed line marks the comet perihelion at 1.79 au, where A(0)f$\rho$ reached a maximum of about 15000 cm, placing K2 among the very active LPC. Before perihelion, A(0)f$\rho$ shows a clear increase that is generally normal as the comet approaches the Sun, reflecting the increased dust release as a result of increased sublimation of ices.

However, a peculiar behavior is observed between -260 and -170 days to perihelion (3.60 to 2.74 au), where Af$\rho$ shows a strange drop followed by a rapid increase, suggesting some correlation with the phase angle. This apparently unusual trend persists even after accounting for the phase angle, suggesting that the phenomenon is intrinsic to the comet. Furthermore, the water production rate increased sharply around 3 au (see Figure \ref{17K2_dust_water_prod}) during which Af$\rho$ and the dust production also started to increase. In particular, it may indicate an exhaustion of CO and/or CO$_2$ ice at least in the near-surface layer, which typically drives sublimation activity between 3 and 6 au.

In any case, it is clear that the comet activity at the time of crossing the snowline underwent a loss of activity, as can be seen with the flattening of the light curve and the stall in the gas production rates. \cite{Kwon2024} observed the same trend using data from the ZTF IRSA archive \footnote{\url{https://irsa.ipac.caltech.edu/applications/ztf/?__action=layout.showDropDown&}} \citep{Masci2019}, covering a range from $\sim$ 14 to $\sim$ 2.3 au. Although the trend is consistent, the Af$\rho$ values differ due to variations in photometry, likely caused by incorrect zero-point magnitude information used in their analysis (\cite{Kwon2024}, personal communication). Post-perihelion, the Af$\rho$ displays a steep decrease with no correlation with the phase angle. This behaviour is usually observed for LPCs.

\subsection{High Resolution Spectroscopy (VLT)}
\subsubsection{Optical spectroscopy with UVES}
\begin{figure}[h]
	\centering	\includegraphics[scale=0.45, trim={1cm 1.5cm 0.15cm 2cm}, clip]{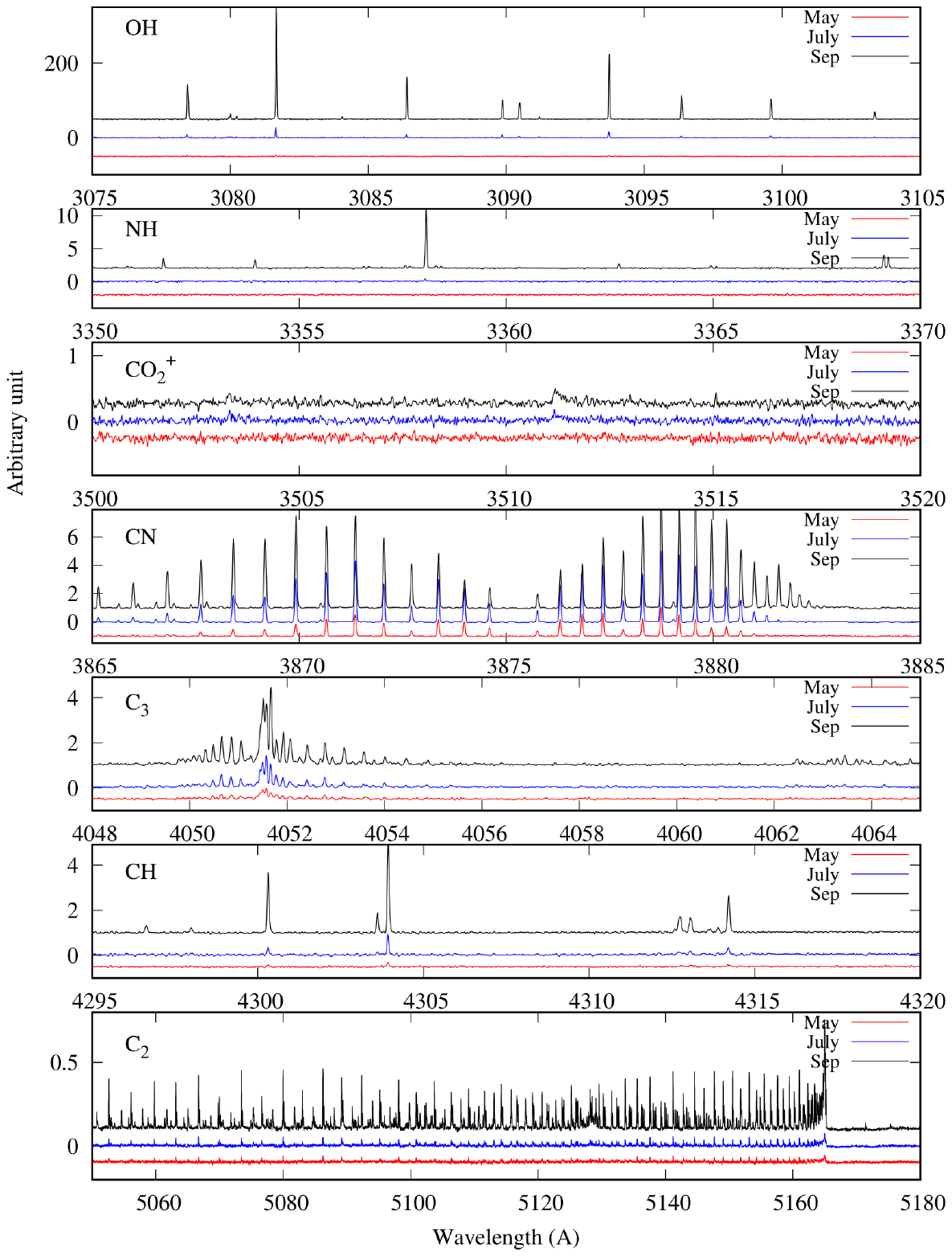}
	\caption{Spectral regions of interest of comet C/2017 K2 acquired with UVES, with detected daughter species in three different epochs: May (${r_h}$=3.23 au), July (${r_h}$=2.73 au), and September (${r_h}$=2.12 au). The flux is reported in arbitrary units for better display.} 
	\label{fig:uves_spectra}
\end{figure}

At the first epoch (May 9, 2022, ${r_h}$=3.23 au), only CN, C$_3$, and C$_2$ are clearly detected with UVES, as in TRAPPIST images. Then at the second epoch (July 5, ${r_h}$=2.73 au), OH and all the usual other species start to be detected (NH, CH, NH$_2$). The emissions became much brighter at the last epoch (Sept 20, ${r_h}$=2.12 au) (see Figure \ref{fig:uves_spectra}). CO$_2^+$ ions are detected, but CO$^+$ ions are never detected, showing that K2 is not a CO$^+$ rich comet like C/2016 R2 observed at the same distance and whose optical spectrum was dominated by CO$^+$ bands \citep{Opitom2019}. But there could also be an observation bias, as we rarely observe CO$^+$ lines in the high-resolution spectra of UVES taken in the last two decades, because the slit is usually placed in the bright inner coma where the bands of the neutral species dominate and can hide the faint CO$^+$ lines. The slit is also tiny (10" in length), and geometrically it might be easy to miss the CO$^+$ thin asymmetric streams (only opposite to the Sun). When the slit is offset from the nucleus in the tail direction, ions can be observed more often. 

The production rates of these daughter species have been derived from a Haser model \citep{Haser1957} using the main emission bands (Table \ref{Qs_UVS}) and compared to expected parent species detected with CRIRES$^+$ simultaneously. The fluxes are, unfortunately, too weak in CN lines to compute the N and C isotopic ratios.

\begin{table}[h!]
\caption{Gas production rates of comet C/2017 K2 from UVES observations}
\label{Qs_UVS}
\centering 
{\renewcommand{\arraystretch}{1.5}
\normalsize
\resizebox{0.5\textwidth}{!}{%
\begin{tabular}{lcccccc}
\hline\hline
Date (2022)   & OH & CN & C$_2$ \\
\hline

May 09  & $3.96 \pm 0.79 \times 10^{27}$ & $7.57 \pm 1.51 \times 10^{25}$ & $2.88 \pm 0.58 \times 10^{25}$  \\

Jul 05 (a)  & $1.23 \pm 0.12 \times 10^{28}$ & $1.07 \pm 0.11 \times 10^{26}$ & $7.03 \pm 1.41 \times 10^{25}$  \\

Jul 05 (b)  & $1.51 \pm 0.16 \times 10^{28}$ & $ 1.15\pm 0.12 \times 10^{26}$ & $ 5.86 \pm 1.17 \times 10^{25}$  \\

Sept 21  & $5.23 \pm 0.51 \times 10^{28}$ & $1.15 \pm 0.12 \times 10^{26}$ & $1.01 \pm 0.20 \times 10^{26}$ \\

Sept 22  & $5.59 \pm 0.57 \times 10^{28}$ & $1.18 \pm 0.11 \times 10^{26}$ & $9.59 \pm 1.92 \times 10^{25}$ \\
\hline
\end{tabular}
}} 
\end{table}

\begin{figure}[h]
	\centering	\includegraphics[width=0.92\linewidth, trim={2cm 1cm 4cm 2cm}]{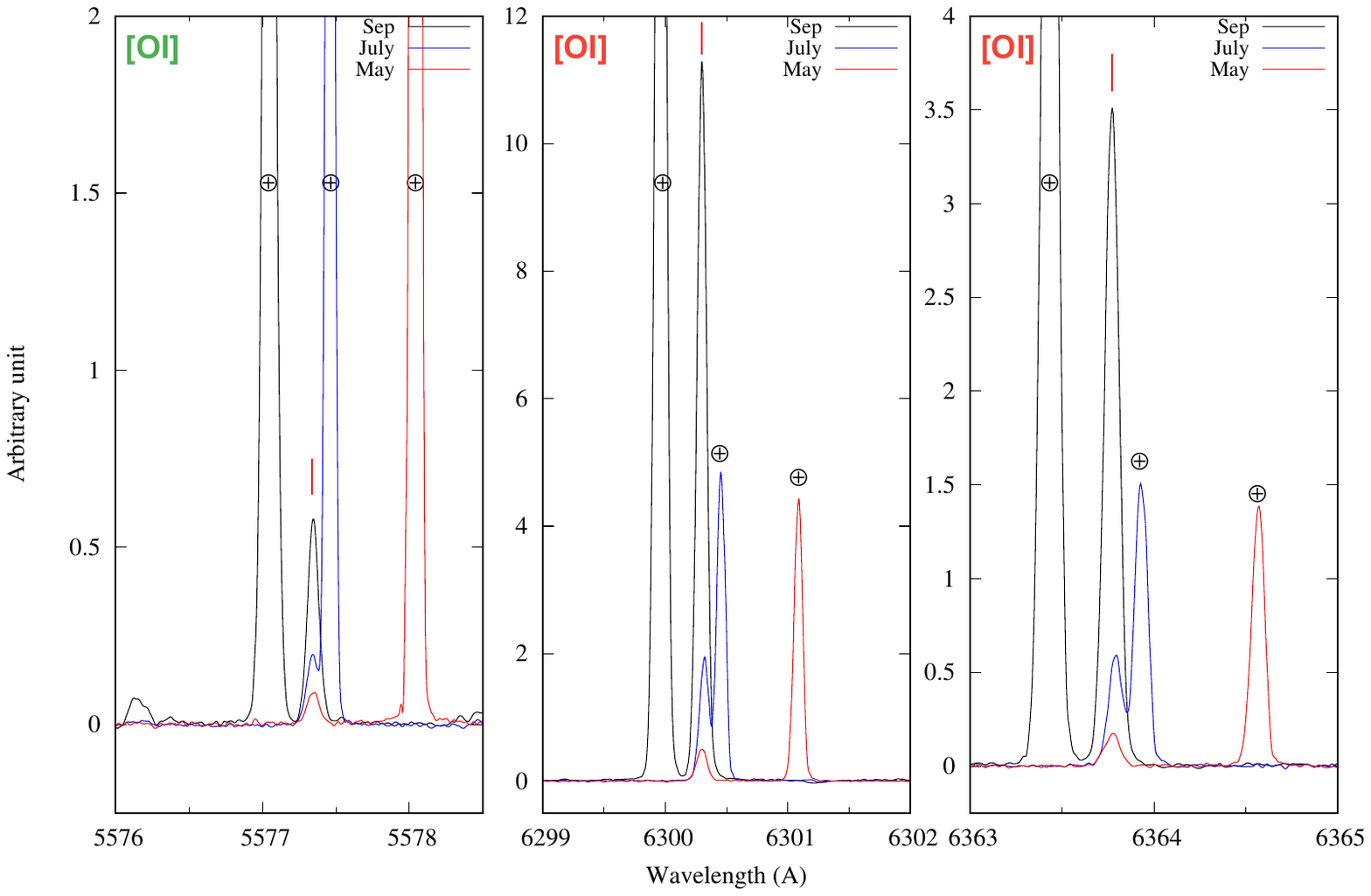}
	\caption{The green forbidden oxygen line [OI] and the red doublet lines detected with UVES at the VLT on the 3 epochs: May 9 (${r_h}$=3.23 au), July 5 (${r_h}$=2.73 au), and Sept 20 (${r_h}$=2.12 au)}.
	\label{fig:OIK2}
\end{figure}

\begin{table}
    \centering
    \caption{Flux measurements and G/R intensity ratios of [OI] emission lines for comet C/2017 K2.}
    \label{G/R_ratios}
    \setlength{\tabcolsep}{6pt} 
    \renewcommand{\arraystretch}{1.5} 
    \resizebox{0.45\textwidth}{!}{ 
    \begin{tabular}{lcccccc}
        \hline
        \hline
        Date (2022) & \multicolumn{3}{c}{Fluxes [ erg/s/cm$^{2}$/A/arcsec$^{2}$]} & G/R \\
                   &($5577 \, \text{\AA}$) &($6300 \, \text{\AA}$) & ($6363 \, \text{\AA}$) &  \\
        \hline
         May 09  & $1.64 \times 10^{-17}$ & $5.76 \times 10^{-17}$ & $1.98 \times 10^{-17}$ & 0.25 $\pm$ 0.013 \\
         Jul 05a & $2.42 \times 10^{-17}$ &$1.52 \times 10^{-16}$ & $4.65 \times 10^{-17}$ & 0.14 $\pm$ 0.007 \\
         Jul 05b & $4.03 \times 10^{-17}$ & $2.43 \times 10^{-16}$ & $7.43 \times 10^{-17}$ & 0.15 $\pm$ 0.008 \\
         Sep 21  & $1.01 \times 10^{-16}$ & $1.18 \times 10^{-15}$ & $3.82 \times 10^{-16}$ & 0.08 $\pm$ 0.004 \\ 
         Sep 22  & $1.02 \times 10^{-16}$ & $1.17 \times 10^{-15}$ & $3.62 \times 10^{-16}$ & 0.08 $\pm$ 0.005\\
        \hline
        \hline
    \end{tabular}}
\end{table}

\begin{figure}[h]
	\centering
	\includegraphics[width=1\linewidth]{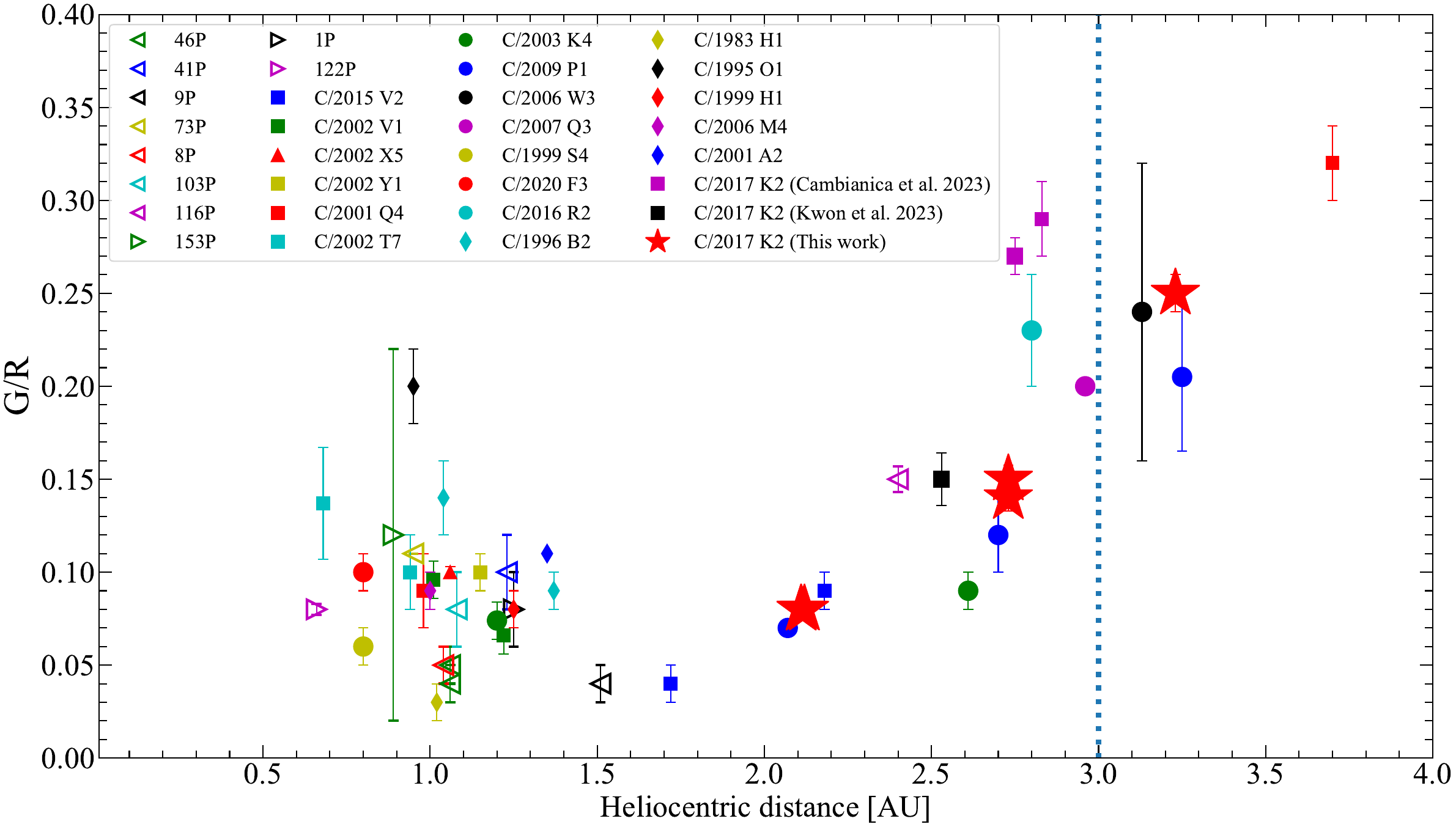}
	\caption{The G/R intensity ratio plotted as a function of the heliocentric distance. The same symbol is used for multiple points of a given comet. Open markers represent short-period comets, and solid markers represent LPCs. The vertical dotted line indicates the distance beyond which water sublimation decreases significantly \citep{Crovisier2000}.}
	\label{fig:G_R_ratios}
\end{figure}

The high resolution and sensitivity of UVES allowed us to detect at each epoch the three [OI] oxygen forbidden lines, the green one at 5577.31 \AA{} and the red doublet at 6300.31 \AA{} and at 6363.78 \AA. Due to the Doppler shift caused by the comet’s velocity with respect to the Earth, the comet lines are well separated from the strong telluric [OI] lines (see Figure \ref{fig:OIK2}). The flux measurements of each line and the so-called green-to-red ratio G/R = I5577 / ( I6300 + I6364 ) \citep{Cochran2001} are reported for each epoch in Table \ref{G/R_ratios}. We observe a clear trend of the ratio with the heliocentric distance from a rather large value of 0.25 above the water sublimation line at 3.0 au \citep{Crovisier2000} down to a value close to 0.1 at 2 au, in excellent agreement with the values reported by \cite{Decock2013}, and references therein with the addition of \cite{Opitom2019}, \cite{Cambianica2021}, \cite{Cambianica2023}, \cite{kwon_k2} and \cite{Aravind2024}, for comets observed at various heliocentric distances (see Figure \ref{fig:G_R_ratios}). This ratio has been commonly used to determine the main parent molecule of the oxygen atoms in the coma, and particularly the relative contribution of the main comet activity drivers' H$_2$O, CO$_2$, and CO, as oxygen is mainly produced by the photodissociation of those species \citep{Festou1981}. The G/R ratio depends indeed on the progenitor, as shown in Table 2 of \cite{Festou1981}, with a value of 0.1 for water and higher values for CO and CO$_2$. Sub-millimeter observations later confirmed the presence of carbon monoxide (CO) in K2’s coma \citep{Yang2021}, and \cite{Cambianica2023} measured a high G/R ratio of 0.28 at 2.8 au. We found a large G/R of 0.25 at 3.2 au but a lower value of 0.15 at 2.7 au, in good agreement with the average value of 0.15 at 2.53 au from \cite{kwon_k2}, using MUSE at the VLT. It then rapidly drops down to 0.08 at 2.1 au, in agreement with the trend of \cite{Decock2013}. This illustrates the quick rise of water sublimation below 3.0 au, confirmed by the UVES data not showing OH lines, a photodissociation product of H$_2$O, until the second epoch. As this G/R trend is followed by several comets at large distances, including K2, we cannot then argue it was especially rich in CO or CO$_2$ compared to others, as an even larger ratio should have been found. The nondetection of CO and the detection of CO$_2^+$ might point to CO$_2$ to be the main contributor at larger heliocentric distances than 2.5 au. It would have been, of course, very interesting to obtain also the ratio at even larger distances than 4.0 au to test this hypothesis.

\begin{figure}[h]
	\centering
	\includegraphics[scale=0.33, trim={0 0.9cm 0 1.75cm}, clip]{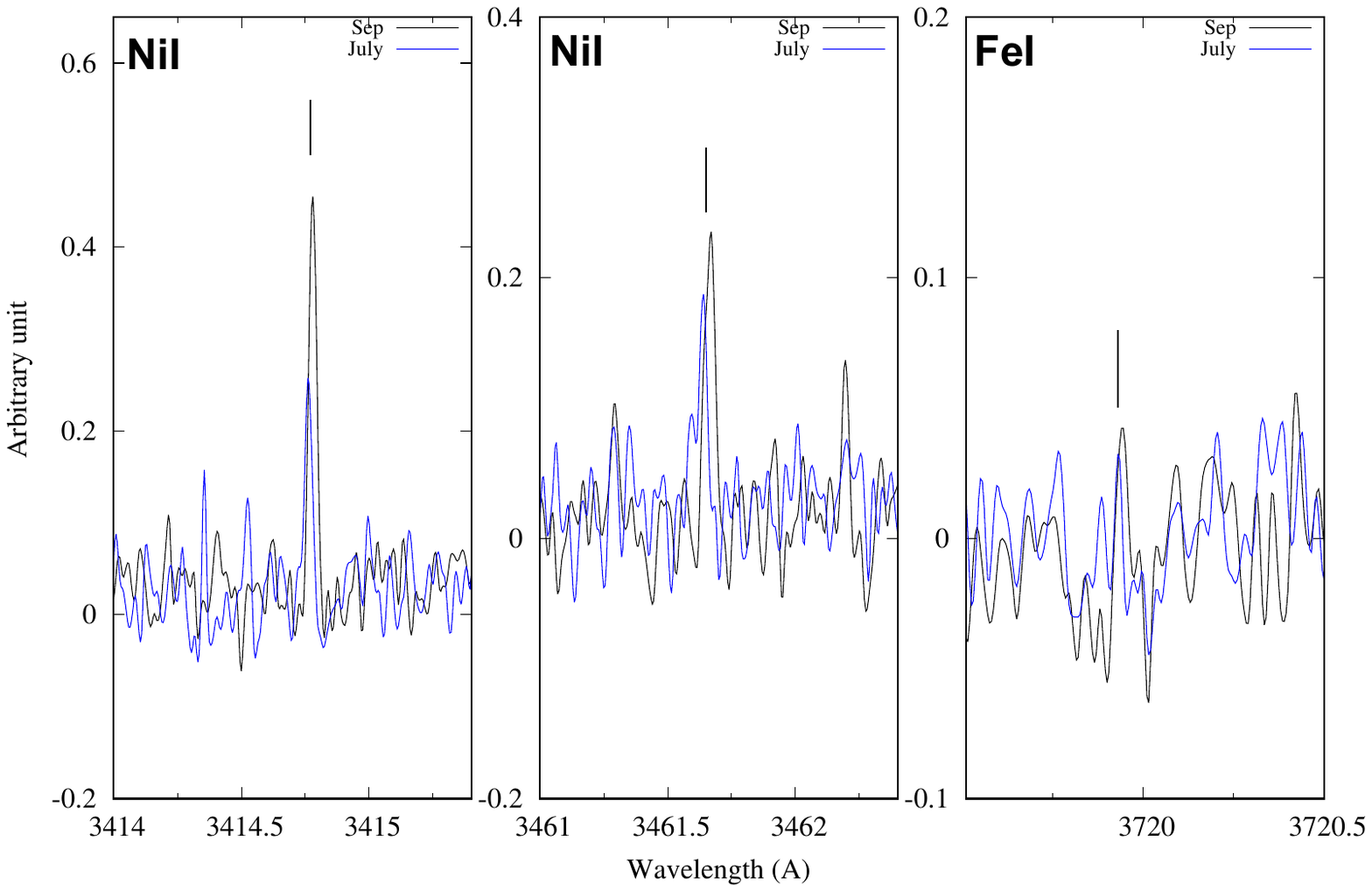}
	\caption{Example of NiI and FeI lines detected in C/2017 K2.}
	\label{fig:NiFeK2}
\end{figure}

\begin{figure}[h]
	\centering
	\includegraphics[width=1\linewidth]{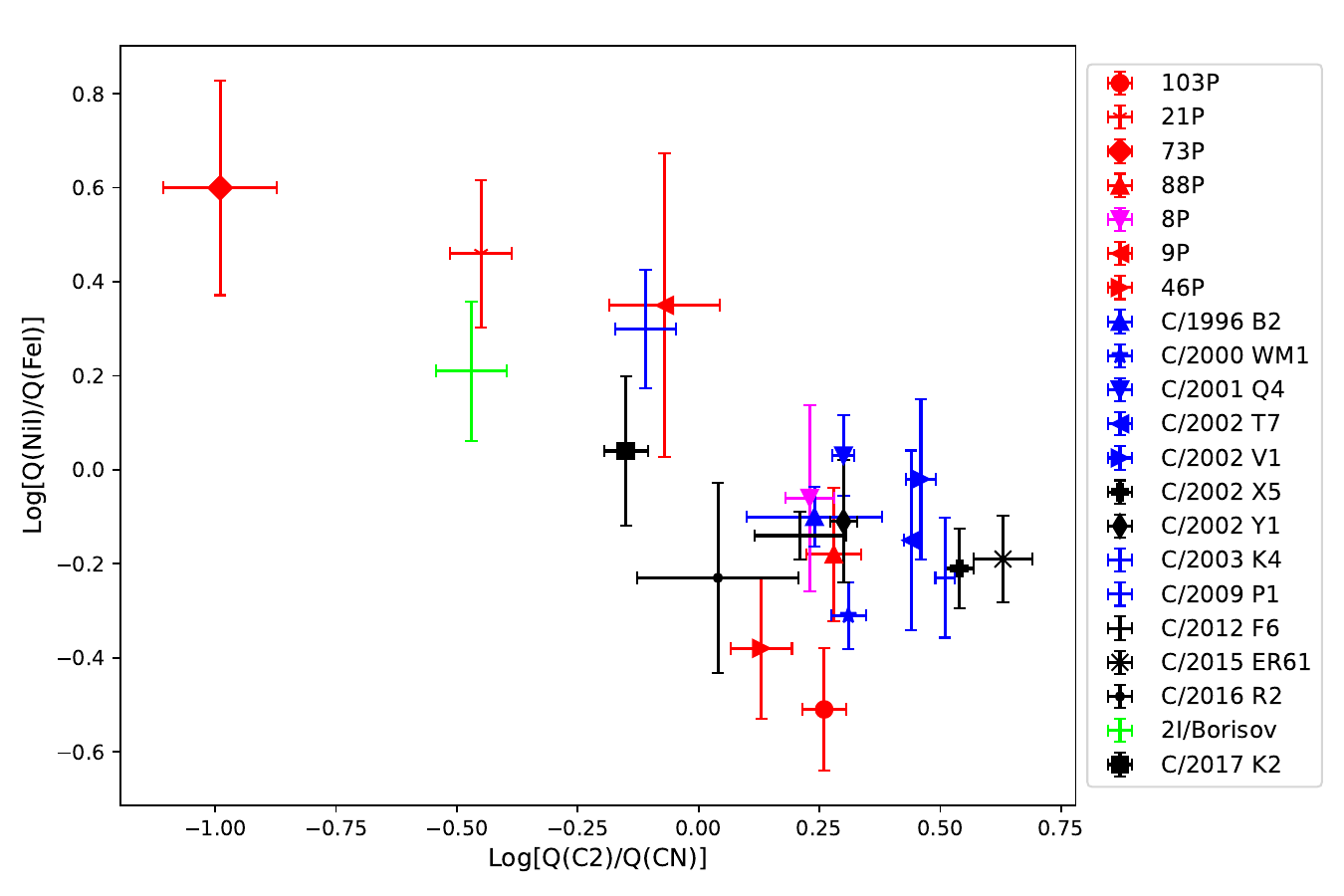}
	\caption{NiI/FeI abundance ratios against the C$_2$/CN  adapted from \cite{Hutsemekers2021}. The colors in the plot represent different dynamical classes: Jupiter-family comets (red), Halley-family comets (pink), long-period comets (blue), and dynamically new comets (black), with a square symbol for comet C/2017 K2.} 
	\label{fig:Ni_Fe_c2_cn_ratios}
\end{figure}

A recent and surprising discovery was the ubiquitous presence of neutral metallic lines of iron and nickel in high-resolution spectra of comets and detected even far from the Sun \citep{Manfroid2021Natur}. In comet K2, we were able to identify and measure the flux of eight nickel lines and two iron lines in the blue spectra of the last two epochs, sometimes with large uncertainties (see Figure \ref{fig:NiFeK2} and Table \ref{NiI_and_FeI_intensity}). We computed the iron and nickel production rates for the last two epochs: log$_{10}$(Q$_{Fe}$) = 22.01 $\pm$ 0.21 (July) and  22.14 $\pm$ 0.21 (Sept) and log$_{10}$(Q$_{Ni}$) = 21.97 $\pm$ 0.04 (July), and 22.23 $\pm$ 0.08 (Sept) using the same model developed in \cite{Manfroid2021Natur} and \cite{Hutsemekers2021}. The ratios for each epochs are then log$_{10}$(Q$_{Ni/Fe}$) = -0.04 $\pm$ 0.22 (July) and 0.09 $\pm$ 0.23 (Sept). They are in good agreement within the error bars with the average value for 17 comets of log$_{10}$(Q$_{Ni/Fe}$) = -0.06 $\pm$ 0.31, but differs by one order of magnitude from the ratio of -1.10 $\pm$ 0.23 estimated in the dust of 1P/Halley \citep{Jessberger1988Natur} and -1.11 $\pm$ 0.09 measured in the coma of the Sun-grazing comet Ikeya-Seki by \cite{Manfroid2021Natur}. Despite being a DNC and active very far from the Sun, the ratio of NiI/FeI in K2 is similar to that of other comets. As shown in Figure \ref{fig:Ni_Fe_c2_cn_ratios}, K2 falls in the middle of the correlation found by \cite{Hutsemekers2021} between the level of Carbon-chain depletion (C$_2$/CN) and the NiI/FeI ratio, of the LPC. The existence of this relation suggests that the diversity of NiI/FeI abundance ratios in comets could be related to the cometary formation rather than to subsequent processes in the coma.

\begin{figure*}[h!]
    \centering
    \includegraphics[width=0.83\linewidth]{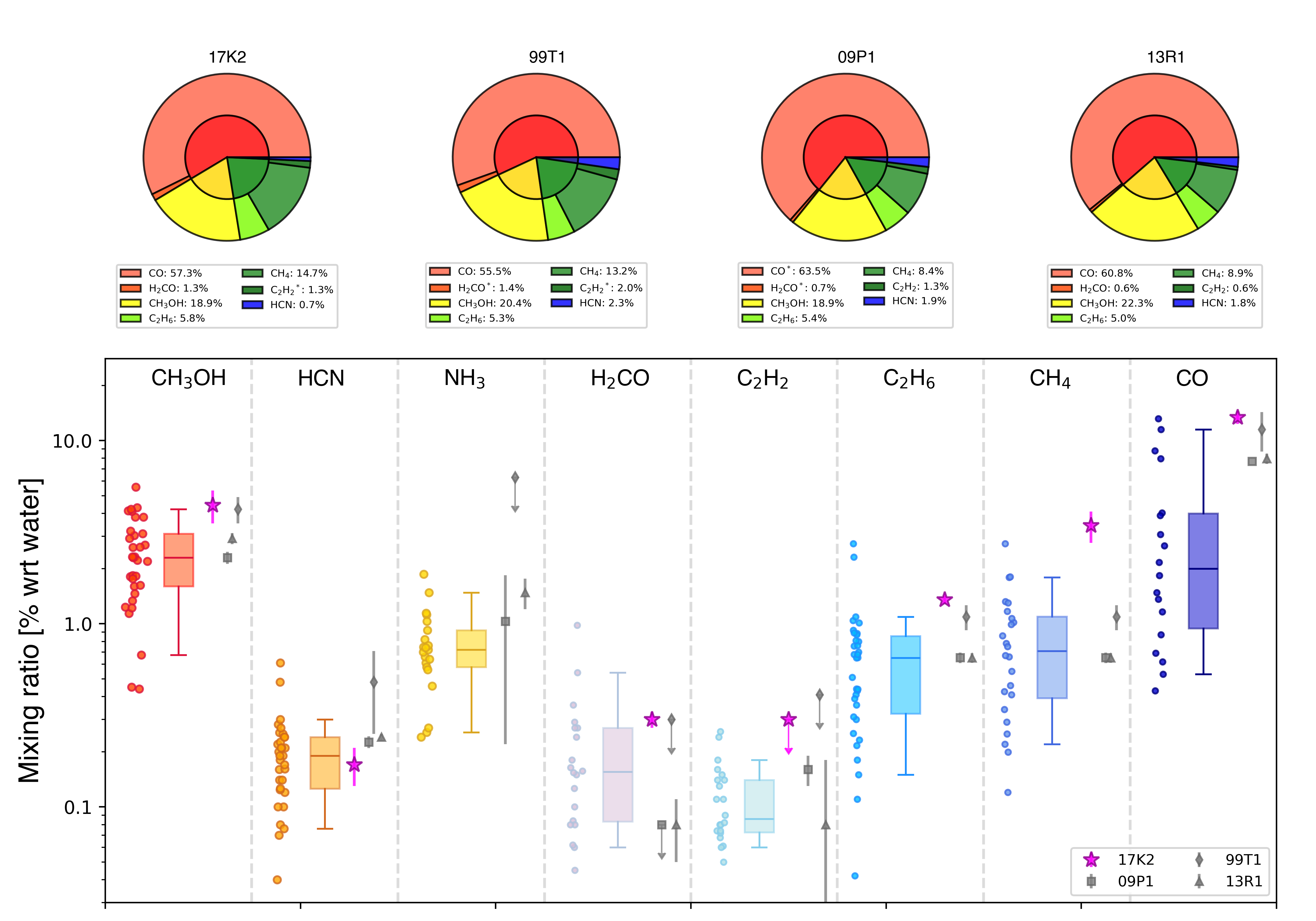}
    \caption{Comparison of K2 with other comets. In the upper panel, the comet is compared with C/2009 P1, C/1999 T1, and C/2013 R1, which show the closest relative abundance proportions. In the bottom plot, for each molecular species, we compare K2 (marked with a magenta star) with the box plot statistic from other infrared results relative to comets observed within 2 au from the Sun \citep{2021AJ....162...74L}. For each box, the middle line corresponds to the median, the box limits to the 25$^{th}$ and 75$^{th}$ percentiles, and the whiskers to the 5$^{th}$ and 95$^{th}$ percentiles. C/2009 P1, C/1999 T1, and C/2013 R1 are also shown with a gray square, diamond, and triangle, respectively.}
    \label{fig:k2_stat}
\end{figure*}

\begin{table}
    \centering
    \caption{Fluxes of NiI and FeI in C/2017 K2.}
    \label{NiI_and_FeI_intensity}
    \setlength{\tabcolsep}{8pt} 
    \renewcommand{\arraystretch}{0.8} 
    \small
    \resizebox{0.48\textwidth}{!}{ 
    \begin{tabular}{ccccc}
        \hline
        \hline
        Line ($\text{\AA}$) & \multicolumn{4}{c}{ Fluxes $\times$ 10$^{-3}$ [ erg/s/cm$^{2}$/A/arcsec$^{2}$ ]} \\
         & July 5 (a) & July 5 (b) & September 21 & September 22 \\
        \hline
        NiI 3414.764   & 8.0 $\pm$ 3.0    & 10.0 $\pm$ 3.0   & 10.0 $\pm$ 3.0    & 17.0 $\pm$ 3.0     \\
        NiI 3446.259   & --               & 4.4 $\pm$ 2.0    & 3.0 $\pm$ 2.0     & 4.7 $\pm$ 3.0      \\
        NiI 3458.460   & --               & --               & --                & 20.0 $\pm$ 10.0    \\
        NiI 3461.652   & 5.0 $\pm$ 2.0    & 5.0 $\pm$ 2.0    & --                & 13.0 $\pm$ 3.0     \\
        NiI 3492.956   & --               & --               & 6.0 $\pm$ 3.0     & --                 \\
        NiI 3515.052   & 3.0 $\pm$ 2.0    & 5.0 $\pm$ 4.0    & 7.9 $\pm$ 3.0     & 11.0 $\pm$ 3.0     \\
        NiI 3524.536   & 6.0 $\pm$ 2.0    & 10.0 $\pm$ 3.0   & 13.7 $\pm$ 3.0    & 15.0 $\pm$ 3.0     \\
        NiI 3619.391   & 4.5 $\pm$ 2.0    & 3.4 $\pm$ 2.0    & 6.2 $\pm$ 3.0     & 2.9 $\pm$ 2.0      \\
        FeI 3719.935  & --               & 3.9 $\pm$ 2.0    & --                & 4.9 $\pm$ 2.0      \\
        FeI 3737.132  & --               & 1.9 $\pm$ 2.0    & --                & 3.1 $\pm$ 2.0      \\
        \hline
        \hline
\end{tabular}}
\end{table}

\begin{table*}[h!]
	\caption{Abundance ratios derived from the optical and infrared data of comet C/2017 K2}
	\label{tab:ir-op_ratios}
	\resizebox{\textwidth}{!}{%
		\begin{tabular}{lcccccccccl}
			\hline
			\hline
			UT Date   & $r_h$ &r$_{\Delta}$  &  \multicolumn{5}{c}{Abundances ratios ($10^{-3}$)}    & Tel/Instrument \\
			&(au) & (au)  & CN/OH & HCN/H$_2$O & C$_2$/OH & C$_2$H$_2$/H$_2$O & C$_2$H$_6$/H$_2$O & \\
			\hline
			2022 Jul 05 &2.73&1.82&--&$<$3.88&--&<4.78&7.73$\pm$2.93& CRIRES$^+$\\
            2022 Jul 07 &2.71&1.82&5.56$\pm$3.89&--&5.98$\pm$4.18&--&--& TRAPPIST\\
            2022 Sep 21 &2.12&2.34&1.68$\pm$0.22&1.48$\pm$0.58&1.76$\pm$0.23&$<$3.14&13.30$\pm$3.20& TRAPPIST $\&$ CRIRES$^+$\\
            2022 Sep 22 &2.11&2.35&2.03$\pm$0.29&2.29$\pm$1.07&1.69$\pm$0.24&$<$4.93&14.50$\pm$3.50& TRAPPIST $\&$ CRIRES$^+$\\
			\hline
			\hline
	\end{tabular}}
\end{table*}

\begin{table} 
  \begin{center}
  \caption{Production rates, and Mixing Ratios (MR, relative to water) retrieved for CRIRES$^+$ observations}
  \label{tab:csmr}
  \renewcommand{\arraystretch}{1.2} 
    \resizebox{0.46\textwidth}{!}{
    \begin{tabular}{ccccc}
    \hline
    \hline
    Date & Molecule  &Q$_{NC}$ &  Q$^{(a)}_{Tot}$ &MR\\
         &          & [$10 ^{26}$ mol/s] &  [$10 ^{26}$ mol/s] &[\% wrw]\\
    \hline
    5 July 2022 & H$_2$O  & 203 $\pm$ 53  &  335 $\pm$ 96 & 100 \\
                & HCN  & $<$ 0.8  & $<$ 1.3  & $<$ 0.4 \\
                & C$_2$H$_2$  &   $<$ 0.9  &  $<$ 1.5 & $<$ 0.5\\
                & C$_2$H$_6$  &    1.57 $\pm$ 0.34 & 2.59 $\pm$ 0.64   & 0.77 $\pm$ 0.32\\
                & CH$_4$  &   7.8 $\pm$ 1.7 &  12.9 $\pm$ 3.2  & 3.8 $\pm$ 0.8\\
                & CO  &    21.8 $\pm$ 5.5  &  36 $\pm$ 10   & 10.7 $\pm$ 2.7  \\
   \hline
    21 Sep 2022 & H$_2$O$^{(b)}$  & 252 $\pm$ 29  & 446 $\pm$ 82  & 100\\
                & CH$_3$OH& 10.5 $\pm$ 3.5 &   19 $\pm$ 7&   4.2 $\pm$ 1.4 \\
                & HCN  &   0.37 $\pm$ 0.12  &  0.66 $\pm$ 0.23  &0.15 $\pm$ 0.05  \\
                & C$_2$H$_2$     &$<$ 0.8 &   $<$ 1.4 & $<$ 0.3\\
                & C$_2$H$_6$     & 3.36 $\pm$ 0.22 & 5.95 $\pm$ 0.92  & 1.33 $\pm$ 0.09\\
                & CH$_4$$^{(b)}$  &    6.5 $\pm$ 3 &  11 $\pm$ 5 & 2.6 $\pm$ 1.2\\
                & CO  &    35 $\pm$ 3 &  60 $\pm$ 10   & 13.9 $\pm$ 1.2  \\
    \hline
    22 Sep 2022 & H$_2$O$^{(b)}$  &   252 $\pm$ 29  & 446 $\pm$ 82  & 100\\
                & CH$_3$OH &11.5 $\pm$ 3.5  &   20 $\pm$ 6&   4.6 $\pm$ 1.2 \\
                & HCN    &0.58 $\pm$ 0.24  &  1.02$\pm$ 0.44  &0.23 $\pm$ 0.10 \\
                & C$_2$H$_2$     &$<$ 1.2  &   $<$ 2.2 & $<$ 0.5\\               
                & C$_2$H$_6$     & 3.64 $\pm$ 0.26 &  6.45 $\pm$ 1.02  & 1.44 $\pm$ 0.10\\
                & CH$_4$$^{(b)}$     & 6.5 $\pm$ 3 &  11 $\pm$ 5 & 2.6 $\pm$ 1.2\\
\hline
\hline
    \end{tabular}}
   \end{center}
   {\footnotesize {\bf Note:} (a) Calculated using a Q$_{scale}$ value of (1.65 $\pm$ 0.20), and (1.77 $\pm$ 0.25), for the July and September observations, respectively; (b) Weighted average of the two observations in September. }	
\end{table}

\begin{table}[]
    \centering 
    \caption{Comparison of total production rates and mixing ratios as retrieved from this work, \cite{2025AJ....169..102E} and \cite{2025arXiv250419849W}.}
    \renewcommand{\arraystretch}{1.2} 
    \resizebox{0.48\textwidth}{!}{
    \begin{tabular}{ccccc}
        \hline
        \hline
          & \multicolumn{2}{c}{CRIRES$^+$}&iSHELL &JWST \\
        r$_h$ [au]  &  2.7 & 2.1  & 2.35 &2.35 \\
         \hline
         &\multicolumn{4}{c}{Total Production Rates [10$^{26}$ mol/s]} \\
         \hline
         H$_2$O&335$\pm$ 96 & 446 $\pm$ 82 &365 $\pm$ 66  &760 $\pm$ 15  \\
         CH$_3$OH& & 19 $\pm$ 7 &17.1 $\pm$ 1.9  &22.2 $\pm$ 0.7  \\
         HCN & $<$ 1.3 & 0.7 $\pm$ 0.2 &1.75 $\pm$ 0.16  &1.7 $\pm$ 0.3  \\
         C$_2$H$_2$& $<$ 1.5 & $<$ 1.4 &1.97 $\pm$ 0.34  &   \\
         C$_2$H$_6$&2.6 $\pm$ 0.6 & 6.3 $\pm$ 0.9 &5.25 $\pm$ 0.67  &2.3 $\pm$ 0.4  \\
         CH$_4$&12.9 $\pm$ 3.2 & 11.0 $\pm$ 5.0 & 15.7 $\pm$ 2.3  &14.8 $\pm$ 0.4  \\
         CO&36 $\pm$ 10 & 60 $\pm$ 10 &63.7 $\pm$ 5.7  &62 $\pm$ 1  \\
         \hline
          &\multicolumn{4}{c}{Mixing Ratios [w.r.t. water]} \\
         \hline
         CH$_3$OH&  & 4.4 $\pm$ 1.2 & 4.69 $\pm$ 1  &2.93 $\pm$ 0.08   \\
         HCN & $<$ 0.4 & 0.2 $\pm$ 0.1 &0.48 $\pm$ 0.10  &0.23 $\pm$ 0.05  \\
         C$_2$H$_2$& $<$ 0.5 & $<$ 0.5 &0.54 $\pm$ 0.13  &  \\
         C$_2$H$_6$&0.77 $\pm$ 0.32 & 1.4 $\pm$ 0.1 & 1.44 $\pm$ 0.32  &0.31 $\pm$ 0.04  \\
         CH$_4$&3.8 $\pm$ 0.8 & 2.6 $\pm$ 1.2 & 4.29 $\pm$ 1.0  &1.95 $\pm$ 0.05  \\
         CO&10.7 $\pm$ 2.7 & 13.9 $\pm$ 1.2 &17.5 $\pm$ 3.5  & 8.2 $\pm$0.3  \\
        \hline
        \hline
    \end{tabular}}
   
    \label{tab:compIR}
\end{table}

\subsubsection{NIR spectroscopy with CRIRES$^+$}
The overall retrieved rotational temperatures, production rates, and mixing ratios (or significant upper limits) obtained from CRIRES$^+$ spectra are reported in Table \ref{tab:csmr}. 

Despite the comet's notable activity even at large heliocentric distances, our infrared observations of K2 reveal a contrasting scenario, characterized by a very weak dust continuum and faint emission lines from many parent species. Because the dust signal in the infrared is a combination of reflected light and thermal emission, the low dust signal we see in our IR spectra could be attributed to the dust being cold since K2 was still quite far away from the Sun, combined with the dust properties (size, composition, etc.). Additional simulations are required to properly explain the differences observed between the optical and the infrared. Considering the volatile component, the infrared counterparts of optically detected species, such as \( \text{H}_2\text{O} \), \( \text{C}_2\text{H}_2 \) and \( \text{HCN} \), were barely discernible, even when the comet appeared bright. This is particularly evident in the spectra taken on May 9, where CO emission lines are at the noise level and too faint to allow a proper estimate of the production rate. In the beginning, given the strong activity observed in the optical, we were expecting spectra dominated by hyper-volatiles such as CO and CH$_4$. Since this was not the case, we assumed that the comet was still too far from the Sun and the observer to be properly sampled in the infrared. Our later analysis and comparison with TRAPPIST and UVES results confirm instead a more complex scenario, where most likely our non-detection could be related to a near-surface depletion of CO before the comet entered the water ice-line (r$_h$ < 2.8 au). 

In Table \ref{tab:compIR}, we compare our results with recent literature outcomes. Our total production rates and mixing ratios are comparable within 2$\sigma$ with those obtained by \cite{2025AJ....169..102E}, with the exception of HCN, for which we get a significantly smaller value. Similarly, our production rates measured in July are consistent with those measured by the James Webb Space Telescope (JWST) when the comet was at 2.35 au from the Sun \citep{2025arXiv250419849W}, even if the latter report a water production rate two times larger than the one we measured, and consequently smaller mixing ratios. 

When compared to other comets, K2 is enriched in almost all the species, and similar to C/1999 T1, C/2009 P1, and C/2013 R1 (see Figure \ref{fig:k2_stat}). A possible scenario that can explain this composition is that the material in this comet formed in a cold, organic-rich environment, where CO was particularly abundant and hydrogenation processes on grains contributed significantly to the abundances of species such as methanol, ethane, and methane. Moreover, as a dynamically new comet, K2 most likely preserved its primordial composition during its storage in the Oort cloud. This pristine and hyper-volatile material has enabled its strong activity even at heliocentric distances larger than 20 au. 

\subsection{Parent and daughter molecules}
To investigate the origin of the radicals observed in the coma of K2, we compared the abundances of the daughter species with the parent molecular species derived from infrared observations using CRIRES$^+$. Table \ref{tab:ir-op_ratios} summarizes the abundances of selected radicals observed in the optical and their potential parent species detected in the infrared during the same period. Our CN/OH ratios align with the HCN/H$_2$O ratios obtained from the infrared observations on approximately the same dates, particularly on September 21, 2022 (see Table \ref{tab:ir-op_ratios}). This indicates that HCN is the primary source of observed CN. The significantly lower C$_2$ abundance compared to C$_2$H$_6$ suggests that C$_2$ is more likely to be dissociated from C$_2$H$_2$ than from C$_2$H$_6$. Previous studies have shown that C$_2$ remains associated with both molecules even at large heliocentric distances, as observed in comet Hale-Bopp \citep{Helbert2005A&A}. Furthermore, no clear parent molecule for C$_3$ has been identified in the K2 infrared spectra, suggesting that C$_3$ may be produced by chemical reactions within the coma.

\section{Summary and conclusion}
\label{summary_conclusion}

In this study, we conducted a comprehensive observational analysis of comet C/2017 K2 (PanSTARRS), a Dynamically New Oort cloud comet that exhibited activity at a large heliocentric distance. Using data from the TRAPPIST telescopes, UVES, and CRIRES$^+$ at the VLT, we characterized the long-term evolution of activity and molecular abundances across multiple wavelengths, while the comet was getting closer to the Sun.
Our TRAPPIST photometric monitoring campaign spanned nearly eight years, covering the comet's approach to perihelion and its post-perihelion activity. Over 271 nights of observation, we acquired more than 2204 broad-band and 174 narrow-band images, providing an extensive dataset to analyze the comet's brightness variations and coma properties. The light curve revealed a steady evolution without significant outbursts, and the observed colors remained about the same over a wide range of heliocentric distances.

The narrow-band photometric data from TRAPPIST and spectroscopic data obtained from UVES and CRIRES$^+$ allowed us to investigate the gas production rates and molecular composition of the coma. CN and C$_2$ radicals were first detected in March 2022, with subsequent detections of other species as the comet moved inward. Production rates exhibited a gradual increase before perihelion, followed by a stabilization phase. The analysis of relative molecular abundances, based on the ratios of C$_2$ and C$_3$ to CN and OH, classified K2 as a "typical" comet, consistent with the taxonomic classification of \cite{A'Hearn1995}. Furthermore, our comparison of parent and daughter species confirmed that HCN is the main source of CN, while C$_2$ is likely to be dissociated from C$_2$H$_2$ rather than C$_2$H$_6$. 

Our findings show that cometary colors do not provide a clear distinction between different dynamical classes, consistent with recent findings by \cite{Carrie2024PSJ}. Notably, comet K2 lies on the bluer end of the color distribution. The temporal evolution of its activity reveals distinct phases linked to the sublimation and exhaustion of the surface of various ices. Our analysis of the variation in absolute magnitude, A(0)f$\rho$ and dust production highlighted a peculiar drop in the brightness and dust production between -260 and -170 days of perihelion. This may indicate variations in the sublimation mechanisms of CO and CO$_2$ relative to H$_2$O. The observed trends suggest that the temporary stall in activity was due to the depletion of hypervolatiles in the near-surface layers, with activity resuming as water sublimation became dominant near perihelion. The switch from a CO$_2$ domination to an H$_2$O dominated coma is also clear from the fast drop in the observed G/R ratio as the comet crossed the water ice sublimation line at $\sim$3 au.

In general, our results provide valuable insight into the long-term evolution of comet K2 and its activity. The findings reinforce the idea that CO and other super-volatile species play a significant role in driving cometary activity at large heliocentric distances. In addition, the compositional characteristics of the comet contribute to a broader understanding of Dynamically New comets and their evolutionary pathways. Furthermore, our study suggests a possible depletion of the outer layers rich in super-volatiles, leading to noticeable changes in the volatile-driven activity as the comet approached the Sun. The existence of a class of comets experiencing progressive depletion of the outermost volatile-rich layers of the nucleus, resulting in fading of activity, will be further investigated in future works. Future observations of similar objects will help refine our understanding of the formation conditions and dynamical history of long-period comets.\\

\begin{acknowledgements}
This publication uses data products from the TRAPPIST project, under the scientific direction of Emmanuel Jehin, Director of Research at the Belgian National Fund for Scientific Research (F.R.S.-FNRS). TRAPPIST-South is funded by F.R.S.-FNRS under grant PDR T.0120.21, and TRAPPIST-North is funded by the University of Liège in collaboration with Cadi Ayyad University of Marrakech. CRIRES$^+$ and UVES results are based on observations under the program 109.23GX, at the European Southern Observatory, Cerro Paranal. S. Hmiddouch acknowledges funding from the Belgian Academy for Research and Higher Education (ARES). M. Lippi acknowledges funding from the “NextGenerationEU” program, in the context of the Italian “Piano Nazionale di Ripresa e Resilienza (PNRR)”, project code SOE\_0000188. M. Vander Donckt acknowledges support from the French-speaking Community of Belgium through its FRIA grant. The authors thank NASA, David Schleicher, and the Lowell Observatory for the loan of a set of HB comet filters.
\end{acknowledgements}

\bibliographystyle{aa}
\bibliography{biblio_17K2}

\begin{appendix}
\onecolumn
\section{Gas production rates and A(0)f$\rho$ parameter from TRAPPIST data.}

\begin{table*}[h!]
    \begin{center}
        \caption{Gas production rates and A(0)f$\rho$ parameter of comet C/2017 K2  (PanSTARRS) from TRAPPIST data. The A(0)f$\rho$ values are computed within a radius aperture of 5-arcsecond from the nucleus and corrected for the phase angle effect.}
        \label{tab:K2Qs}
        \resizebox{1\textwidth}{!}{%
            \setlength{\tabcolsep}{4pt}
            \begin{tabular}{lcccccccccccc}
                \hline
                \hline
                UT Date & $r_h$ & r$_{\Delta}$ & \multicolumn{5}{c}{Production rates (molec/s) }   &  \multicolumn{4}{c}{A(0)f$\rho$ (cm)}\\
                &(au) & (au)  & OH &NH &CN & C$_2$ & C$_3$ & BC & RC & GC& Rc & \\
                \hline
2017-10-25 & 15.18 & 15.31 &   &   &   &   &   &   &   &   & 3773.40$\pm$717.30  \\
2018-11-06 & 12.89 & 13.08 &   &   &   &   &   &   &   &   & 5986.88$\pm$562.23  \\
2018-11-10 & 12.87 & 13.06 &   &   &   &   &   &   &   &   & 5123.75$\pm$409.26  \\
2019-04-08 & 11.92 & 11.85 &   &   &   &   &   &   &   &   & 6809.18$\pm$249.87  \\
2019-05-10 & 11.71 & 11.56 &   &   &   &   &   &   &   &   & 6739.79$\pm$212.81  \\
2019-05-28 & 11.59 & 11.41 &   &   &   &   &   &   &   &   & 5911.20$\pm$256.14  \\
2019-06-05 & 11.54 & 11.35 &   &   &   &   &   &   &   &   & 6225.22$\pm$323.87  \\
2019-06-09 & 11.51 & 11.33 &   &   &   &   &   &   &   &   & 6903.24$\pm$222.92  \\
2019-06-11 & 11.50 & 11.31 &   &   &   &   &   &   &   &   & 6693.24$\pm$260.36 \\
2019-06-12 & 11.49 & 11.31 &   &   &   &   &   &   &   &   & 6121.04$\pm$248.04 \\
2019-06-25 & 11.40 & 11.22 &   &   &   &   &   &   &   &   & 6826.96$\pm$204.70  \\
2019-07-17 & 11.26 & 11.11 &   &   &   &   &   &   &   &   & 6489.91$\pm$617.77  \\
2019-07-20 & 11.24 & 11.10 &   &   &   &   &   &   &   &   & 6881.36$\pm$577.68  \\
2019-07-22 & 11.23 & 11.09 &   &   &   &   &   &   &   &   & 7358.43$\pm$375.48  \\
2019-07-28 & 11.18 & 11.06 &   &   &   &   &   &   &   &   & 6669.67$\pm$230.02  \\
2019-09-26 & 10.78 & 10.89 &   &   &   &   &   &   &   &   & 6730.76$\pm$249.43  \\
2019-10-03 & 10.73 & 10.87 &   &   &   &   &   &   &   &   & 7157.57$\pm$260.69  \\
2019-10-23 & 10.60 & 10.81 &   &   &   &   &   &   &   &   & 7109.72$\pm$207.93  \\
2019-10-24 & 10.59 & 10.81 &   &   &   &   &   &   &   &   & 7260.67$\pm$229.85  \\
2019-11-03 & 10.52 & 10.77 &   &   &   &   &   &   &   &   & 7110.86$\pm$293.56  \\
2020-01-27 & 9.94 & 10.18 &   &   &   &   &   &   &   &   & 7385.39$\pm$256.92  \\
2020-01-28 & 9.93 & 10.17 &   &   &   &   &   &   &   &   & 7304.06$\pm$318.19  \\
2020-02-04 & 9.88 & 10.10 &   &   &   &   &   &   &   &   & 7118.60$\pm$203.10  \\
2020-02-08 & 9.85 & 10.05 &   &   &   &   &   &   &   &   & 7101.09$\pm$340.09  \\
2020-03-10 & 9.63 & 9.69 &   &   &   &   &   &   &   &   & 7832.26$\pm$762.57 \\
2020-03-17 & 9.59 & 9.61 &   &   &   &   &   &   &   &   & 8064.85$\pm$252.32 \\
2020-04-03 & 9.46 & 9.40 &   &   &   &   &   &   &   &   & 8108.65$\pm$182.73  \\
2020-04-15 & 9.38 & 9.25 &   &   &   &   &   &   &   &   & 7519.40$\pm$263.56  \\
2020-04-17 & 9.37 & 9.23 &   &   &   &   &   &   &   &   & 7946.26$\pm$147.37  \\
2020-04-27 & 9.29 & 9.12 &   &   &   &   &   &   &   &   & 7994.35$\pm$167.97  \\
2020-05-05 & 9.24 & 9.03 &   &   &   &   &   &   &   &   & 7308.59$\pm$207.45  \\
2020-05-09 & 9.21 & 8.98 &   &   &   &   &   &   &   &   & 8162.83$\pm$220.96  \\
2020-05-20 & 9.13 & 8.88 &   &   &   &   &   &   &   &   & 7793.10$\pm$152.64 \\
2020-05-24 & 9.10 & 8.84 &   &   &   &   &   &   &   &   & 7740.78$\pm$120.90  \\
2020-06-02 & 9.04 & 8.76 &   &   &   &   &   &   &   &   & 7714.21$\pm$143.66  \\
2020-07-11 & 8.75 & 8.49 &   &   &   &   &   &   &   &   & 7550.97$\pm$249.67 \\
2020-07-23 & 8.66 & 8.44 &   &   &   &   &   &   &   &   & 7506.43$\pm$89.91 \\
2020-07-24 & 8.65 & 8.43 &   &   &   &   &   &   &   &   & 7793.29$\pm$122.18 \\
2020-08-05 & 8.56 & 8.39 &   &   &   &   &   &   &   &   & 7863.07$\pm$265.63 \\
2020-08-12 & 8.51 & 8.38 &   &   &   &   &   &   &   &   & 8013.76$\pm$95.72 \\
2020-08-27 & 8.40 & 8.35 &   &   &   &   &   &   &   &   & 7692.02$\pm$154.72  \\
2020-09-07 & 8.31 & 8.33 &   &   &   &   &   &   &   &   & 8771.15$\pm$99.06  \\
2020-09-21 & 8.21 & 8.32 &   &   &   &   &   &   &   &   & 8503.51$\pm$204.74  \\
2020-10-05 & 8.11 & 8.30 &   &   &   &   &   &   &   &   & 9216.04$\pm$115.88  \\
2020-10-11 & 8.06 & 8.29 &   &   &   &   &   &   &   &   & 8868.92$\pm$173.21  \\
2020-10-24 & 7.96 & 8.26 &   &   &   &   &   &   &   &   & 9022.92$\pm$161.57  \\
2020-12-07 & 7.63 & 8.08 &   &   &   &   &   &   &   &   & 9439.39$\pm$324.59  \\
2021-04-22 & 6.55 & 6.31 &   &   &   &   &   &   &   &   & 9637.61$\pm$234.19  \\
2021-05-06 & 6.44 & 6.11 &   &   &   &   &   &   &   &   & 10034.74$\pm$118.95  \\
2021-05-12 & 6.39 & 6.04 &   &   &   &   &   &   &   &   & 10363.67$\pm$80.99  \\
2021-06-05 & 6.19 & 5.73 &   &   &   &   &   &   &   &   & 10333.07$\pm$63.64  \\
2021-06-22 & 6.06 & 5.58 &   &   &   &   &   &   &   &   & 11414.00$\pm$106.92  \\
2021-06-28 & 6.01 & 5.54 &   &   &   &   &   &   &   &   & 11205.46$\pm$98.23  \\
2021-09-09 & 5.38 & 5.40 &   &   &   &   &   &   &   &   & 12457.29$\pm$87.67  \\
2021-09-13 & 5.35 & 5.40 &   &   &   &   &   &   &   &   & 11995.34$\pm$99.31  \\
2022-01-12 & 4.29 & 5.00 &   &   &   &   &   &   &   &   & 12140.12$\pm$176.53  \\
2022-01-13 & 4.29 & 4.98 &   &   &   &   &   &   &   &   & 11988.58$\pm$175.16  \\
2022-02-02 & 4.11 & 4.69 &   &   &   &   &   &   &   &   & 12096.66$\pm$113.28  \\
2022-03-13 & 3.75 & 3.94 &   &   &   &   &   &   &   &   & 12256.02$\pm$102.55  \\
2022-03-25 & 3.64 & 3.67 &   &   &   &   &   &   &   &   & 11632.88$\pm$185.45  \\
2022-03-28 & 3.62 & 3.60 &   &   &1.27$\pm$0.10e+26 & 1.49$\pm$0.25e+26 &   &   &   &   &   \\
2022-03-29 & 3.61 & 3.58 &   &   &   &   &   &   &   &   & 12237.24$\pm$268.79 \\
2022-03-31 & 3.59 & 3.53 &   &   & 1.35$\pm$0.10e+26 & 1.22$\pm$0.51e+26 &   &   &   &   & 11261.88$\pm$285.32 \\
2022-04-01 & 3.58 & 3.51 &   &   &   &   &   &   &   &   & 11879.42$\pm$130.81 \\
2022-04-02 & 3.57 & 3.48 &   &   &   &   &   &   &   &   & 11398.86$\pm$87.10 \\
2022-04-07 & 3.52 & 3.37 &   &   & 1.45$\pm$0.10e+26 & 9.69$\pm$1.33e+25 &   &   &   &   & 11291.50$\pm$128.80  \\
2022-04-08 & 3.52 & 3.34 &   &   & 1.48$\pm$0.11e+26 & 1.57$\pm$0.18e+26 &   &   &   &   & 10897.88$\pm$218.57  \\
2022-04-09 & 3.51 & 3.32 &   &   & 1.43$\pm$0.09e+26 & 1.61$\pm$0.26e+26 &   & 10399.28$\pm$181.51 & 12016.04$\pm$269.77 &   & 10996.01$\pm$284.99 \\
2022-04-13 & 3.47 & 3.23 &   &   & 1.54$\pm$0.10e+26 & 1.05$\pm$0.40e+26 &   & 9684.19$\pm$186.76 & 11590.88$\pm$233.66 & 10186.83$\pm$169.10 & 10533.13$\pm$199.45 \\
2022-04-27 & 3.34 & 2.90 & 1.66$\pm$2.67e+28 &   & 1.64$\pm$0.12e+26 & 1.17$\pm$0.18e+26 &   & 9196.37$\pm$150.20 & 11067.33$\pm$108.19 & 9736.40$\pm$106.83 & 10056.05$\pm$206.61\\
    \end{tabular}}
    \end{center}
\end{table*}

\newpage
\begin{table*}
    \begin{center}
        \resizebox{1\textwidth}{!}{%
            \setlength{\tabcolsep}{4pt}
            \begin{tabular}{lcccccccccccc}
                UT Date & $r_h$ & r$_{\Delta}$ & \multicolumn{5}{c}{Production rates (molec/s) }   &  \multicolumn{4}{c}{A(0)f$\rho$ (cm)} \\
                &(au) & (au)  & OH &NH &CN & C$_2$ & C$_3$ & BC & RC & GC& Rc & \\
                \hline

2022-04-28 & 3.33 & 2.88 &   &   & 1.65$\pm$0.17e+26 & 1.61$\pm$0.14e+26 & 4.43$\pm$0.42e+25 & 9538.22$\pm$159.89 & 11321.08$\pm$77.96 & 9914.76$\pm$81.62 & 10297.28$\pm$87.87  \\
2022-05-03 & 3.29 & 2.77 & 1.55$\pm$2.75e+28 &   & 1.84$\pm$0.10e+26 & 1.25$\pm$0.19e+26 &   & 9519.65$\pm$189.15 & 10717.32$\pm$102.76 & 9706.76$\pm$193.03 & 10458.82$\pm$68.31 \\
2022-05-10 & 3.23 & 2.61 &   &   &   &   &   & 9969.13$\pm$225.41 & 11222.31$\pm$75.99 & 9454.35$\pm$153.78 & 10208.94$\pm$64.32 \\
2022-05-16 & 3.17 & 2.49 & 1.73$\pm$1.04e+28 &   & 1.78$\pm$0.17e+26 & 1.54$\pm$0.17e+26 & 4.41$\pm$0.46e+25 & 9421.87$\pm$273.13 & 10642.21$\pm$250.59 & 9352.73$\pm$263.42 & 10170.12$\pm$80.43 \\
2022-05-24 & 3.10 & 2.33 & 1.71$\pm$2.54e+28 &   &   &   & 4.27$\pm$0.36e+25 & 9290.15$\pm$113.26 & 10757.36$\pm$62.29 & 9352.46$\pm$67.50 & 9896.96$\pm$80.55  \\
2022-05-25 & 3.09 & 2.31 & 1.73$\pm$2.53e+28 &   &   &   & 2.96$\pm$0.23e+25 &   &   &   &    \\
2022-05-27 & 3.07 & 2.27 &   &   & 2.03$\pm$0.11e+26 &   & 3.25$\pm$0.56e+25 & 9504.32$\pm$144.52 & 10969.99$\pm$52.51 & 9427.29$\pm$98.70 & 9958.57$\pm$82.59  \\
2022-06-08 & 2.97 & 2.08 & 1.77$\pm$0.85e+28 &   &   &   & 3.62$\pm$0.34e+25 &   &   &   &   \\
2022-06-09 & 2.96 & 2.06 & 1.82$\pm$0.76e+28 &   & 2.07$\pm$0.11e+26 & 1.66$\pm$0.11e+26 & 3.94$\pm$0.33e+25 & 8403.57$\pm$154.27 & 9984.15$\pm$56.12 & 8737.93$\pm$59.34 & 9257.98$\pm$45.74  \\
2022-06-11 & 2.94 & 2.04 &   &   &   &   &   &   &   &   & 9622.98$\pm$55.25  \\
2022-06-13 & 2.92 & 2.01 & 2.15$\pm$2.31e+28 &   & 2.09$\pm$0.10e+26 & 1.72$\pm$0.17e+26 &   & 8353.38$\pm$144.28 & 10330.02$\pm$77.70 & 9062.00$\pm$79.13 & 9644.26$\pm$73.33 \\
2022-06-18 & 2.88 & 1.95 & 2.64$\pm$0.58e+28 &   &   &   & 3.93$\pm$0.23e+25 &   &   &   &   \\
2022-06-29 & 2.78 & 1.85 & 3.21$\pm$2.57e+28 &   & 1.98$\pm$0.10e+26 &   & 4.23$\pm$0.28e+25 & 8160.92$\pm$116.02 & 9976.38$\pm$98.02 & 8800.84$\pm$60.67 &   \\
2022-06-30 & 2.77 & 1.85 &   &   & 2.04$\pm$0.09e+26 & 2.08$\pm$0.11e+26 & 5.65$\pm$0.27e+25 & 8559.22$\pm$94.85 & 10329.01$\pm$59.58 & 8938.91$\pm$55.58 & 9485.21$\pm$32.65 \\
2022-07-04 & 2.74 & 1.83 &   &   & 2.08$\pm$0.10e+26 & 2.14$\pm$0.10e+26 & 4.97$\pm$0.27e+25 & 8699.99$\pm$110.97 & 10578.42$\pm$42.53 & 9188.18$\pm$51.66 & 9424.96$\pm$37.03 \\
2022-07-07 & 2.71 & 1.82 & 3.58$\pm$2.50e+28 &   & 1.99$\pm$0.09e+26 &   &   &   & 9760.89$\pm$41.91 &   &  \\
2022-08-17 & 2.37 & 2.00 & 6.70$\pm$0.65e+28 & 6.99$\pm$3.12e+25 & 2.08$\pm$0.09e+26 & 2.12$\pm$0.12e+26 & 4.63$\pm$0.35e+25 & 9354.06$\pm$234.69 & 11412.33$\pm$68.11 &   &   \\
2022-08-18 & 2.36 & 2.01 & 7.77$\pm$0.79e+28 &   &   & 2.08$\pm$0.10e+26 & 4.42$\pm$0.31e+25 &   &   &   &   \\
2022-08-19 & 2.35 & 2.02 & 6.39$\pm$0.62e+28 & 1.24$\pm$0.30e+26 & 1.98$\pm$0.06e+26 & 2.24$\pm$0.07e+26 &   &   &   & 9895.09$\pm$54.92 & 10516.60$\pm$66.31 \\
2022-08-21 & 2.34 & 2.04 & 7.28$\pm$0.67e+28 &   & 1.76$\pm$0.05e+26 &   &   &   &   &   & 10293.35$\pm$33.15  \\
2022-08-22 & 2.34 & 2.04 &   &   &   &   &   & 8756.62$\pm$79.16 & 11317.90$\pm$29.44 & 9474.49$\pm$46.04 & \\
2022-08-23 & 2.33 & 2.05 & 7.76$\pm$0.79e+28 &   & 1.84$\pm$0.06e+26 & 1.74$\pm$0.07e+26 & 4.08$\pm$0.18e+25 &   & 11244.40$\pm$33.43 & 9495.22$\pm$56.85 & 10373.68$\pm$43.67\\
2022-08-24 & 2.32 & 2.06 & 7.72$\pm$0.80e+28 &   & 1.78$\pm$0.06e+26 &   &   & 9093.19$\pm$85.92 & 11494.16$\pm$38.14 & 9694.42$\pm$63.94 & 10489.86$\pm$60.71 \\
2022-08-25 & 2.32 & 2.07 & 6.40$\pm$0.68e+28 &   &   &   &   &   & 11055.06$\pm$33.76 &   & 10541.86$\pm$33.13 \\
2022-08-27 & 2.30 & 2.09 & 7.77$\pm$0.94e+28 &   & 1.97$\pm$0.07e+26 & 1.84$\pm$0.10e+26 & 4.29$\pm$0.22e+25 & 9108.89$\pm$101.31 & 11525.30$\pm$35.50 & 9624.97$\pm$77.85 & 10390.12$\pm$50.79 \\
2022-08-29 & 2.29 & 2.11 & 7.92$\pm$0.89e+28 & 1.01$\pm$0.22e+26 & 1.94$\pm$0.07e+26 & 1.86$\pm$0.10e+26 &   & 8788.41$\pm$101.57 & 11209.29$\pm$39.64 & 9398.84$\pm$70.91 & 10404.27$\pm$55.99  \\
2022-08-31 & 2.26 & 2.14 & 8.17$\pm$0.79e+28 &   & 1.89$\pm$0.07e+26 &   &   &   &   &   & 10404.26$\pm$40.15 \\
2022-09-01 & 2.26 & 2.14 & 7.95$\pm$1.12e+28 &   &   &   &   &   & 11454.41$\pm$37.48 & 9470.84$\pm$56.70 & 10415.31$\pm$40.58 \\
2022-09-02 & 2.26 & 2.15 & 9.08$\pm$1.12e+28 &   & 2.04$\pm$0.08e+26 & 1.84$\pm$0.12e+26 & 4.27$\pm$0.27e+25 & 8871.96$\pm$146.23 & 10985.71$\pm$72.02 &   & 10303.84$\pm$63.37 \\
2022-09-04 & 2.23 & 2.18 & 8.42$\pm$0.80e+28 &   & 1.88$\pm$0.06e+26 &   &   &   &   &   & 10665.87$\pm$48.86 \\
2022-09-05 & 2.23 & 2.18 &   &   &   &   & 3.68$\pm$0.22e+25 & 9060.62$\pm$111.16 &   &   &   \\
2022-09-06 & 2.23 & 2.19 &   & 1.64$\pm$0.53e+26 & 2.03$\pm$0.12e+26 &   &   &   &   &   &   \\
2022-09-10 & 2.20 & 2.23 & 9.13$\pm$1.08e+28 &   & 1.95$\pm$0.10e+26 & 1.55$\pm$0.10e+26 & 4.36$\pm$0.27e+25 & 9129.88$\pm$155.01 & 11443.29$\pm$41.76 & 9874.95$\pm$74.51 & 10462.40$\pm$60.76 \\
2022-09-14 & 2.17 & 2.28 & 8.80$\pm$0.91e+28 &   &   &   &   &   &   &   & 10537.60$\pm$52.00 \\
2022-09-15 & 2.17 & 2.28 &   &   & 1.83$\pm$0.07e+26 & 1.50$\pm$0.09e+26 &   & 9149.68$\pm$104.65 & 11114.51$\pm$49.07 &   & \\
2022-09-18 & 2.14 & 2.32 & 7.96$\pm$0.97e+28 & 1.13$\pm$0.36e+26 &   &   &   &   &   &   &  \\
2022-09-19 & 2.14 & 2.32 &   &   & 1.74$\pm$0.07e+26 & 1.65$\pm$0.09e+26 & 4.02$\pm$0.23e+25 & 8613.29$\pm$137.74 &   &   &  \\
2022-09-20 & 2.12 & 2.33 & 9.26$\pm$1.18e+28 & 1.36$\pm$0.31e+26 &   &   &   &   &   &   &  \\
2022-09-21 & 2.12 & 2.34 & 1.03$\pm$0.13e+29 & 1.16$\pm$0.35e+26 & 1.73$\pm$0.07e+26 & 1.81$\pm$0.08e+26 & 3.46$\pm$0.23e+25 & 8778.92$\pm$125.11 & 11023.20$\pm$45.71 &   &  \\
2022-09-22 & 2.12 & 2.35 &   &   & 1.91$\pm$0.12e+26 & 1.59$\pm$0.09e+26 & 4.01$\pm$0.28e+25 & 9186.00$\pm$126.76 & 11420.02$\pm$53.39 & 9573.12$\pm$83.00 & 10765.38$\pm$66.64 \\
2022-09-23 & 2.10 & 2.36 & 9.43$\pm$1.23e+28 & 1.17$\pm$0.30e+26 &   &   &   &   &   &   & 10721.47$\pm$59.64 \\
2022-09-24 & 2.10 & 2.36 &   &   & 1.78$\pm$0.07e+26 & 1.82$\pm$0.08e+26 & 3.98$\pm$0.29e+25 & 9027.18$\pm$137.67 & 11104.90$\pm$71.01 & 9649.48$\pm$88.50 &  \\
2022-09-26 & 2.09 & 2.39 & 8.06$\pm$1.01e+28 & 3.39$\pm$0.26e+26 &   &   &   &   &   &   & \\
2022-09-30 & 2.06 & 2.42 & 1.08$\pm$0.15e+29 &   &   &   &   &   &   &   & 10977.33$\pm$60.48 \\
2022-10-01 & 2.06 & 2.43 & 1.11$\pm$0.16e+29 &   & 1.98$\pm$0.08e+26 & 2.23$\pm$0.10e+26 & 4.65$\pm$0.32e+25 & 8952.98$\pm$149.53 & 11829.10$\pm$73.71 &   & 11104.32$\pm$67.78\\
2022-10-02 & 2.06 & 2.43 &   &   & 1.98$\pm$0.08e+26 & 1.63$\pm$0.11e+26 & 4.34$\pm$0.34e+25 & 9257.92$\pm$157.17 & 11781.13$\pm$59.43 &   & 10995.86$\pm$77.05 \\
2022-10-06 & 2.03 & 2.46 & 1.14$\pm$0.18e+29 &   &   &   &   &   &   &   & 11219.20$\pm$76.82\\
2022-10-07 & 2.03 & 2.46 &   &   & 2.04$\pm$0.09e+26 & 2.04$\pm$0.11e+26 &   & 9065.49$\pm$153.03 & 11713.52$\pm$60.41 &   & \\
2022-10-12 & 1.99 & 2.50 & 1.08$\pm$0.18e+29 &   &   &   &   &   &   &   & 11120.46$\pm$101.29\\
2022-10-13 & 1.99 & 2.50 &   &   & 1.96$\pm$0.10e+26 & 1.98$\pm$0.12e+26 &   & 9798.04$\pm$194.68 & 11829.96$\pm$81.96 &   & \\
2022-10-14 & 1.98 & 2.51 & 1.18$\pm$0.22e+29 &   &   &   &   &   &   &   &  \\
2022-10-15 & 1.98 & 2.52 & 1.20$\pm$0.23e+29 &   & 2.14$\pm$0.11e+26 & 2.13$\pm$0.13e+26 &   & 10377.08$\pm$197.29 & 12643.32$\pm$58.22 &   & 11639.50$\pm$130.00\\
2022-10-16 & 1.98 & 2.52 &   &   & 2.01$\pm$0.10e+26 & 2.25$\pm$0.13e+26 &   & 10272.49$\pm$197.62 & 12408.85$\pm$102.82 &   & 11470.76$\pm$96.15\\
2022-10-19 & 1.96 & 2.54 & 1.22$\pm$0.23e+29 &   &   &   &   &   &   &   & \\
2022-10-20 & 1.95 & 2.54 &   & 3.09$\pm$0.61e+26 & 2.01$\pm$0.10e+26 & 2.01$\pm$0.11e+26 &   &   &   &   & \\
2022-10-21 & 1.95 & 2.54 &   &   &   & 2.68$\pm$0.12e+26 & 5.67$\pm$0.37e+25 &   &   &   & \\
2022-10-22 & 1.94 & 2.55 &   & 3.87$\pm$0.70e+26 &   &   &   &   &   &   &  \\
2022-10-23 & 1.94 & 2.56 &   & 4.18$\pm$0.73e+26 &   & 3.06$\pm$0.14e+26 & 6.21$\pm$0.41e+25 &   &   &   & \\
2022-10-24 & 1.94 & 2.56 &   &   &   & 3.20$\pm$0.15e+26 & 6.19$\pm$0.42e+25 &   &   &   & \\
2023-01-28 & 1.87 & 2.26 & 1.33$\pm$0.29e+29 &   & 2.84$\pm$0.14e+26 & 2.61$\pm$0.17e+26 &   & 13528.06$\pm$259.55 & 14924.44$\pm$63.84 & 13337.70$\pm$133.27 & 14127.19$\pm$102.92\\
2023-01-29 & 1.87 & 2.26 & 1.41$\pm$0.30e+29 &   & 2.82$\pm$0.14e+26 &   & 6.82$\pm$0.52e+25 & 13436.79$\pm$256.50 &   &   & 13814.90$\pm$98.71\\
2023-01-30 & 1.87 & 2.25 & 1.37$\pm$0.28e+29 &   & 2.69$\pm$0.13e+26 &   & 6.93$\pm$0.53e+25 & 13409.13$\pm$243.76 &   &   & 13484.26$\pm$95.29\\
2023-01-31 & 1.88 & 2.25 & 1.22$\pm$0.25e+29 &   & 2.60$\pm$0.13e+26 &   & 7.07$\pm$0.50e+25 & 12725.15$\pm$233.41 &   &   & 12759.80$\pm$88.69\\
2023-02-23 & 1.98 & 2.24 & 9.92$\pm$1.45e+28 & 1.69$\pm$0.38e+26 & 2.22$\pm$0.09e+26 & 2.18$\pm$0.10e+26 & 5.38$\pm$0.33e+25 & 10774.56$\pm$169.30 & 11938.66$\pm$40.10 & 10613.67$\pm$97.15 & 11343.67$\pm$70.31\\
2023-02-26 & 2.00 & 2.25 & 1.29$\pm$0.30e+29 & 3.57$\pm$0.78e+26 & 2.49$\pm$0.18e+26 & 2.32$\pm$0.15e+26 &   &   &   &   & 11419.44$\pm$67.93\\
2023-02-27 & 2.00 & 2.25 &   &   &   &   & 5.66$\pm$0.33e+25 & 11001.79$\pm$144.41 & 12708.20$\pm$46.60 & 10853.59$\pm$77.64 & 11854.54$\pm$66.27\\
2023-03-04 & 2.03 & 2.27 & 9.94$\pm$1.38e+28 & 2.43$\pm$0.51e+26 & 2.25$\pm$0.10e+26 &   & 5.40$\pm$0.41e+25 & 10955.75$\pm$186.81 & 11938.53$\pm$79.65 & 10635.30$\pm$116.84 & 11488.29$\pm$85.20\\
2023-03-13 & 2.09 & 2.34 & 7.16$\pm$0.99e+28 &   &   &   &   &   &   &   &  \\
2023-03-14 & 2.09 & 2.34 &   & 2.04$\pm$0.47e+26 & 2.22$\pm$0.09e+26 & 2.11$\pm$0.08e+26 & 5.58$\pm$0.33e+25 & 9935.40$\pm$159.32 &   &   & 11085.31$\pm$75.40\\
2023-03-20 & 2.13 & 2.40 & 7.10$\pm$0.96e+28 & 1.26$\pm$0.59e+26 &   &   &   &   &   &   & \\
2023-03-21 & 2.13 & 2.40 &   &   & 2.06$\pm$0.09e+26 & 1.74$\pm$0.08e+26 & 4.89$\pm$0.30e+25 & 9539.89$\pm$176.97 & 10538.70$\pm$48.97 & 9343.81$\pm$84.38 &\\
2023-03-23 & 2.15 & 2.43 &   &   & 2.19$\pm$0.10e+26 &   &   &   &   &   & \\
2023-03-24 & 2.15 & 2.43 &   &   &   & 1.73$\pm$0.09e+26 & 5.55$\pm$0.33e+25 & 9418.33$\pm$122.62 &   &   & 10562.88$\pm$55.61\\
2023-03-26 & 2.18 & 2.46 & 7.83$\pm$1.11e+28 &   &   &   &   &   &   &   & \\
2023-03-27 & 2.18 & 2.46 &   &   & 2.27$\pm$0.09e+26 & 1.74$\pm$0.10e+26 & 5.96$\pm$0.37e+25 & 9463.55$\pm$152.08 &   & 9645.06$\pm$89.73 & \\
            \end{tabular}}
    \end{center}
\end{table*}

\newpage
\begin{table*}
    \begin{center}
        \resizebox{1\textwidth}{!}{%
            \setlength{\tabcolsep}{4pt}
            \begin{tabular}{lcccccccccccc}
                UT Date & $r_h$ & r$_{\Delta}$ & \multicolumn{5}{c}{Production rates (molec/s) }   &  \multicolumn{4}{c}{A(0)f$\rho$ (cm)}\\
                &(au) & (au)  & OH &NH &CN & C$_2$ & C$_3$ & BC & RC & GC& Rc & \\
                \hline
2023-03-30 & 2.20 & 2.51 & 8.11$\pm$1.40e+28 &   &   &   &   &   &   &   & \\
2023-03-31 & 2.20 & 2.51 &   & 2.33$\pm$0.71e+26 & 2.30$\pm$0.10e+26 & 1.53$\pm$0.10e+26 & 5.49$\pm$0.44e+25 &   &   &   & \\
2023-04-10 & 2.28 & 2.65 & 3.87$\pm$0.73e+28 &   & 2.17$\pm$0.10e+26 &   &   &   &   &   & \\
2023-04-11 & 2.28 & 2.65 &   &   &   & 1.56$\pm$0.10e+26 &   & 9781.91$\pm$176.84 &   &   & \\
2023-04-18 & 2.34 & 2.76 &   &   & 2.08$\pm$0.08e+26 & 1.32$\pm$0.10e+26 & 4.88$\pm$0.37e+25 &   &   &   & 9480.55$\pm$67.10\\
2023-04-19 & 2.35 & 2.77 & 5.33$\pm$1.21e+28 &   &   &   &   & 10010.17$\pm$172.54 & 10451.85$\pm$54.06 &   & 10070.73$\pm$50.00\\
2023-04-26 & 2.41 & 2.88 &   &   &   &   &   &   &   &   & 9537.65$\pm$56.30 \\
2023-04-28 & 2.42 & 2.91 &   &   &   &   &   &   &   &   & 10318.65$\pm$65.31 \\
2023-04-29 & 2.43 & 2.92 &   &   &   &   &   &   &   &   & 9696.57$\pm$67.18 \\
2023-05-03 & 2.46 & 2.98 &   &   &   &   &   &   &   &   & 9338.61$\pm$63.24\\
2023-05-07 & 2.49 & 3.04 &   &   &   &   &   &   &   &   & 9433.18$\pm$61.48 \\
2023-08-25 & 3.46 & 3.92 &   &   &   &   &   &   &   &   & 9724.36$\pm$50.56\\
2023-09-16 & 3.65 & 3.84 &   &   &   &   &   &   &   &   & 9228.83$\pm$61.84 \\
2023-09-20 & 3.69 & 3.82 & 3.51$\pm$53.70e+26 &   & 2.02$\pm$0.25e+25 &   &   & 6664.21$\pm$163.02 & 7101.00$\pm$83.87 &   & 7387.04$\pm$118.02\\
2023-09-24 & 3.73 & 3.80 &   &   &   &   &   &   &   &   & 8428.02$\pm$56.18 \\
2023-09-25 & 3.73 & 3.80 &   &   &   & 1.02$\pm$0.15e+26 &   & 6782.46$\pm$262.86 & 7289.63$\pm$75.36 &   & 8787.31$\pm$45.49\\
2023-09-29 & 3.77 & 3.77 &   &   &   &   &   &   &   &   & 8617.09$\pm$68.03\\
2023-10-05 & 3.83 & 3.74 &   &   & 2.04$\pm$0.68e+25 & 8.78$\pm$1.94e+25 & 7.73$\pm$2.41e+24 &   &   &   & 7403.56$\pm$65.80\\
2023-10-06 & 3.84 & 3.73 &   &   &   &   &   &   &   &   & 7364.98$\pm$63.78 \\
2023-10-08 & 3.85 & 3.72 &   &   &   &   &   &   &   &   & 7378.67$\pm$44.75\\
2023-10-10 & 3.87 & 3.71 &   &   &   &   &   &   &   &   & 7505.57$\pm$65.87 \\
2023-10-11 & 3.88 & 3.70 &   &   & 2.97$\pm$0.37e+25 &   &   &   &   &   & 7077.14$\pm$77.66\\
2023-10-15 & 3.92 & 3.67 &   &   &   &   &   &   &   &   & 7178.48$\pm$41.37 \\
2023-10-20 & 3.96 & 3.65 &   &   &   &   &   &   &   &   & 7523.33$\pm$36.77 \\
2023-10-26 & 4.01 & 3.61 &   &   &   &   &   & 6313.46$\pm$150.46 & 6873.33$\pm$97.94 &   & 7180.85$\pm$76.53\\
2023-11-08 & 4.13 & 3.55 &   &   &   &   &   &   &   &   & 7308.62$\pm$52.94\\
2023-11-12 & 4.17 & 3.54 &   &   &   &   &   &   &   &   & 6716.60$\pm$64.68\\
2023-11-14 & 4.19 & 3.53 &   &   &   &   &   &   &   &   & 6694.86$\pm$53.90 \\
2023-11-16 & 4.20 & 3.52 &   &   &   &   &   &   &   &   & 7426.78$\pm$85.88 \\
2023-11-18 & 4.22 & 3.52 &   &   &   &   &   &   &   &   & 7480.30$\pm$69.18 \\
2023-11-22 & 4.26 & 3.52 &   &   &   &   &   &   &   &   & 6906.71$\pm$50.66 \\
2023-12-06 & 4.38 & 3.53 &   &   &   &   &   &   &   &   & 6947.59$\pm$36.70 \\
2023-12-07 & 4.39 & 3.54 &   &   &   &   &   &   &   &   & 6187.96$\pm$49.97  \\
2023-12-09 & 4.42 & 3.55 &   &   &   &   &   &   &   &   & 6828.59$\pm$55.06  \\
2023-12-10 & 4.43 & 3.55 &   &   &   &   &   &   &   &   & 6934.24$\pm$62.69  \\
2023-12-11 & 4.43 & 3.56 &   &   &   &   &   &   &   &   & 6101.98$\pm$89.04 \\
2023-12-14 & 4.46 & 3.58 &   &   &   &   &   &   &   &   & 6812.69$\pm$66.66  \\
2023-12-18 & 4.50 & 3.61 &   &   &   &   &   &   &   &   & 6726.98$\pm$60.82  \\
2023-12-21 & 4.52 & 3.63 &   &   &   &   &   &   &   &   & 5994.54$\pm$55.49  \\
2024-01-02 & 4.63 & 3.76 &   &   &   &   &   &   &   &   & 6460.88$\pm$51.61  \\
2024-01-06 & 4.66 & 3.81 &   &   &   &   &   &   &   &   & 6656.26$\pm$31.48  \\
2024-01-23 & 4.82 & 4.11 &   &   &   &   &   &   &   &   & 6503.41$\pm$73.57  \\
2024-01-24 & 4.82 & 4.13 &   &   &   &   &   &   &   &   & 6291.64$\pm$77.35  \\
2024-01-27 & 4.85 & 4.19 &   &   &   &   &   &   &   &   & 5967.72$\pm$59.84  \\
2024-02-03 & 4.91 & 4.34 &   &   &   &   &   &   &   &   & 6087.57$\pm$51.16 \\
2024-02-11 & 4.98 & 4.52 &   &   &   &   &   &   &   &   & 6413.08$\pm$93.37  \\
2024-02-12 & 4.99 & 4.54 &   &   &   &   &   &   &   &   & 6198.75$\pm$54.54  \\
2024-02-21 & 5.07 & 4.75 &   &   &   &   &   &   &   &   & 6251.93$\pm$117.50  \\
2024-02-29 & 5.14 & 4.95 &   &   &   &   &   &   &   &   & 6261.90$\pm$73.92  \\
2024-03-04 & 5.17 & 5.05 &   &   &   &   &   &   &   &   & 5418.19$\pm$66.65  \\
2024-03-10 & 5.22 & 5.18 &   &   &   &   &   &   &   &   & 6076.09$\pm$40.21  \\
2024-04-03 & 5.42 & 5.77 &   &   &   &   &   &   &   &   & 5622.44$\pm$51.66  \\
2024-04-09 & 5.47 & 5.91 &   &   &   &   &   &   &   &   & 5344.47$\pm$67.99  \\
2024-04-11 & 5.50 & 5.97 &   &   &   &   &   &   &   &   & 5408.12$\pm$103.88  \\
2024-09-12 & 6.76 & 6.98 &   &   &   &   &   &   &   &   & 4619.09$\pm$156.38  \\
2024-09-14 & 6.78 & 6.96 &   &   &   &   &   &   &   &   & 4548.82$\pm$178.25  \\
2024-09-26 & 6.88 & 6.86 &   &   &   &   &   &   &   &   & 4648.64$\pm$189.65  \\
2024-09-27 & 6.88 & 6.85 &   &   &   &   &   &   &   &   & 4745.24$\pm$197.46  \\
2024-10-04 & 6.94 & 6.79 &   &   &   &   &   &   &   &   & 4517.08$\pm$81.08  \\
2024-10-30 & 7.15 & 6.57 &   &   &   &   &   &   &   &   & 3865.80$\pm$56.82  \\
2024-11-04 & 7.18 & 6.54 &   &   &   &   &   &   &   &   & 3705.50$\pm$143.13  \\
2024-11-11 & 7.24 & 6.50 &   &   &   &   &   &   &   &   & 3125.83$\pm$130.47  \\
2024-11-23 & 7.33 & 6.47 &   &   &   &   &   &   &   &   & 3807.00$\pm$95.94 \\
2024-11-27 & 7.37 & 6.47 &   &   &   &   &   &   &   &   & 3051.94$\pm$107.26 \\
2024-12-03 & 7.42 & 6.48 &   &   &   &   &   &   &   &   & 3737.69$\pm$112.79  \\
2024-12-11 & 7.47 & 6.50 &   &   &   &   &   &   &   &   & 4024.36$\pm$201.05  \\
2024-12-19 & 7.54 & 6.55 &   &   &   &   &   &   &   &   & 2846.36$\pm$237.66  \\
2025-01-04 & 7.66 & 6.72 &   &   &   &   &   &   &   &   & 2718.69$\pm$127.23  \\
2025-01-20 & 7.78 & 6.96 &   &   &   &   &   &   &   &   & 4196.35$\pm$114.38 \\
2025-01-25 & 7.82 & 7.05 &   &   &   &   &   &   &   &   & 3892.18$\pm$190.63  \\
2025-02-01 & 7.88 & 7.21 &   &   &   &   &   &   &   &   & 3880.53$\pm$219.32  \\
2025-02-17 & 8.00 & 7.57 &   &   &   &   &   &   &   &   & 3693.66$\pm$117.04  \\
2025-03-26 & 8.28 & 8.47 &   &   &   &   &   &   &   &   & 3475.26$\pm$262.88  \\
2025-04-20 & 8.46 & 9.04 &   &   &   &   &   &   &   &   & 2680.62$\pm$329.86 \\

                \hline
                \hline
            \end{tabular}}
    \end{center}
\end{table*}

\onecolumn
\section{Light curve, absolute magnitude evolution, and g-factors.}

\begin{figure}[h]
	\centering	\includegraphics[width=0.76\linewidth]{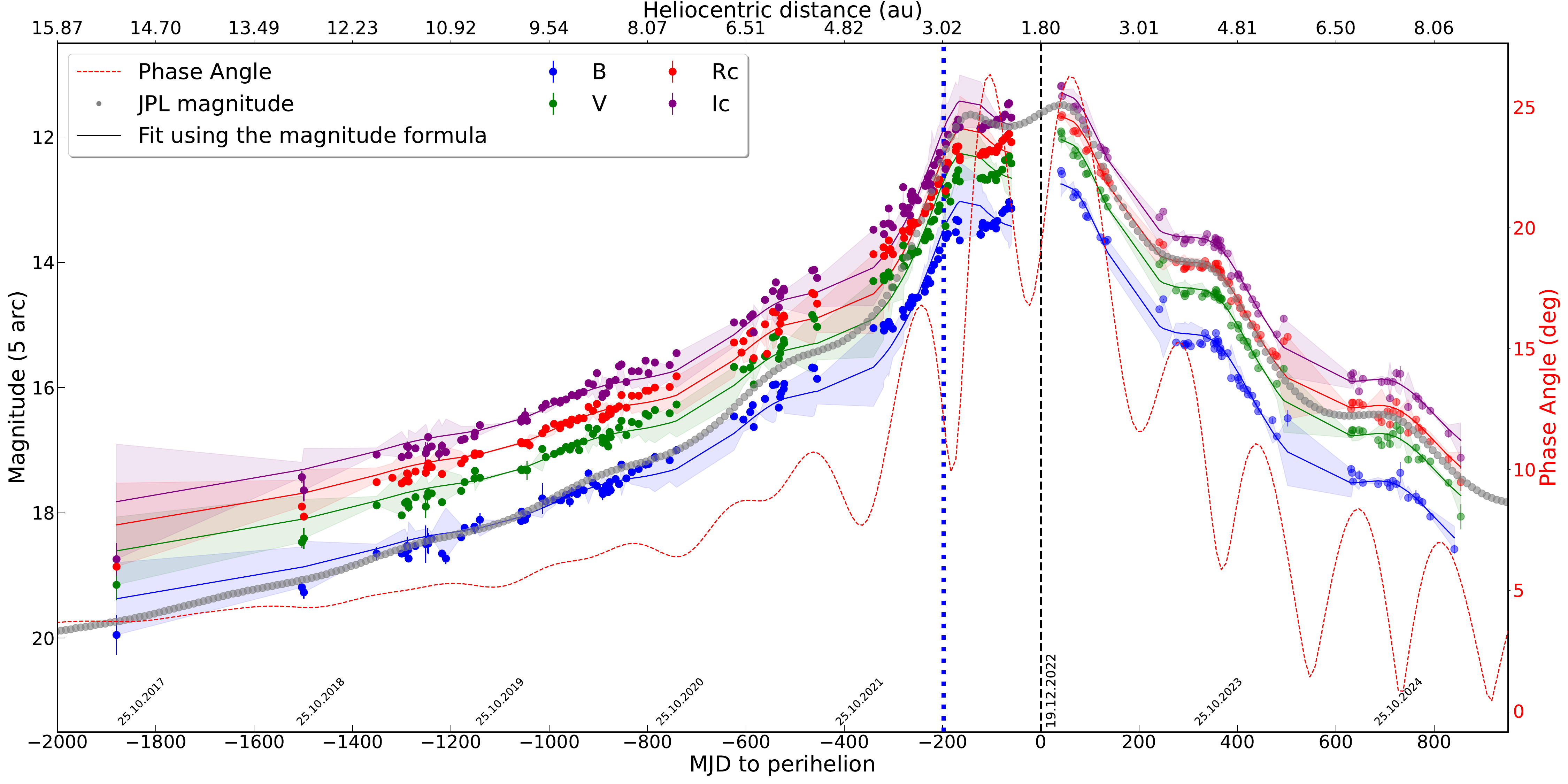}
	\caption{TRAPPIST Light curve of comet C/2017 K2 measured within a radius aperture of 5-arcseconds as a function of time and distance to perihelion. The vertical dashed line indicates the perihelion at 1.79 au on December 19, 2022. The vertical blue dotted line represents the water ice sublimation boundary ($\sim$ 3 au).} 
	\label{fig:mag_K2}
\end{figure}

\begin{figure}[h]
	\centering
	\includegraphics[width=0.75\linewidth]{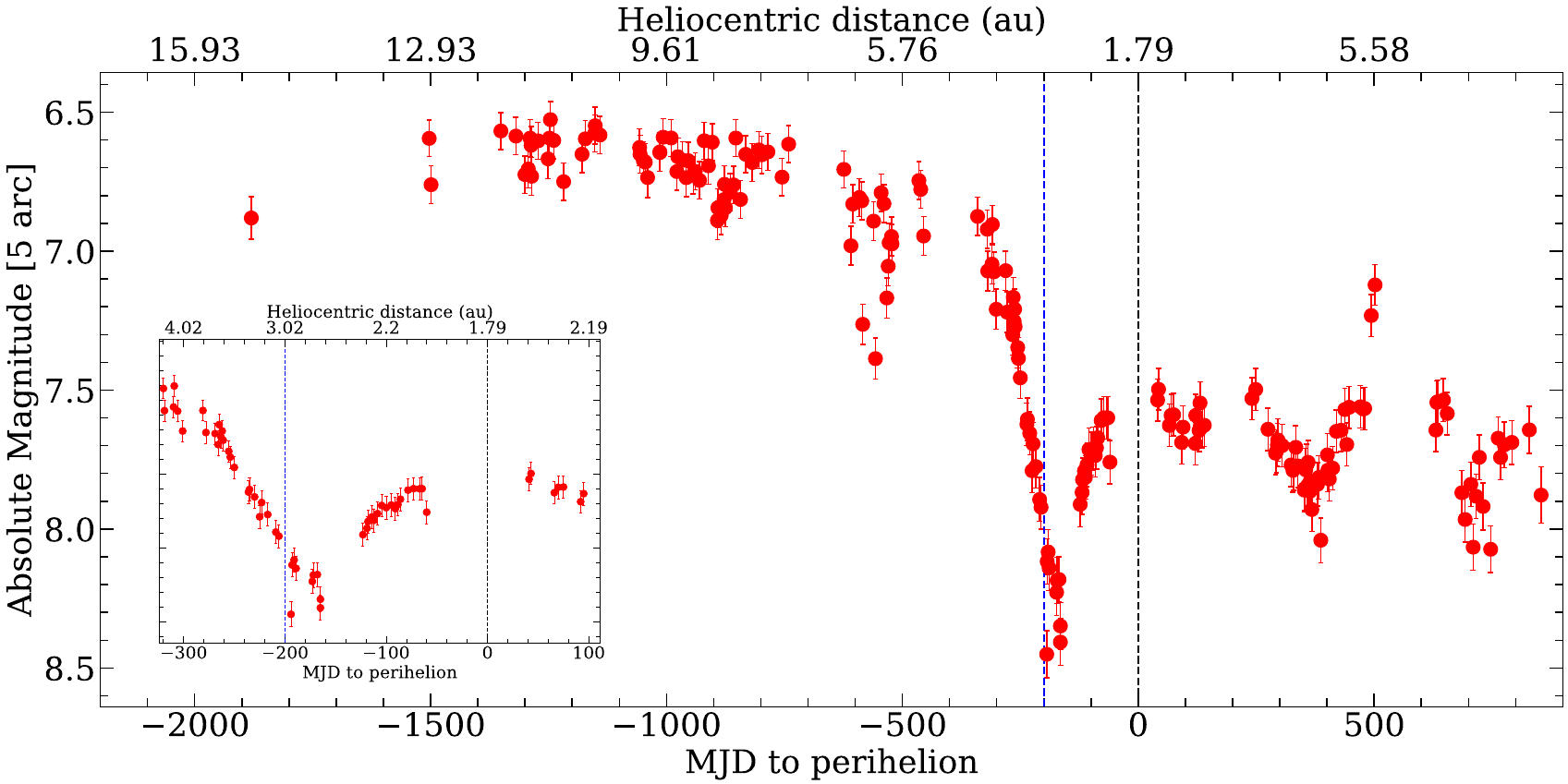}
	\caption{The absolute magnitude of comet C/2017 K2 as a function of days to perihelion. The vertical blue dotted line represents the water ice sublimation boundary ($\sim$ 3 au) \citep{womack2017, Crovisier2000}, and the dashed line represents the perihelion. The inset is a zoomed-in region of the light curve around 3 au.}
	\label{fig:absmag_K2}
\end{figure}

\begin{table}[h!]
        \begin{center}
                \caption{The scale lengths, lifetimes, and the fluorescence efficiencies of the different radicals at 1 au scaled by r$_h^{-2}$. The fluorescence efficiencies are taken from Schleicher's website.}
                \begin{tabular}{lcccc} 
                        \hline
                        Molecules  &  Parent   & Daughter  & Lifetime  & g-factors \\
                                   &  (km)     &  (km)     &(s)  & erg s$^{-1}$ mol$^{-1}$\\
                        \hline 
                        OH(0,0) &  2.4 $\times$10$^4$ & 1.6 $\times$10$^5$ & 1.6 $\times$10$^5$& 1.49$\times$10$^{-15}$\\
                        NH(0,0) &  5.0 $\times$10$^4$ & 1.5 $\times$10$^5$ &  1.3 $\times$10$^4$ & 6.27$\times$10$^{-14}$\\
                        CN($\Delta\upsilon$=0) &  1.3 $\times$10$^4$ & 2.1 $\times$10$^5$ & 2.1 $\times$10$^5$ & 2.62$\times$10$^{-13}$ \\
                        C$_3$($\lambda$=4050 {\AA}{}) &  2.8 $\times$10$^3$ & 2.7 $\times$10$^5$ & 2.7 $\times$10$^5$ & 1.00$\times$10$^{-12}$ \\
                        C$_2$($\Delta\upsilon$=0) & 2.2 $\times$10$^4$ & 6.6 $\times$10$^4$ & 6.6 $\times$10$^4$ &4.50$\times$10$^{-13}$ \\
                        \hline  
                        \hline
                        \label{g-factors_table} 
                \end{tabular}
        \end{center}
\end{table}

\onecolumn
\section{CRIRES$^+$ spectra.}
\begin{figure*}[h!]
    \centering	
    \includegraphics[scale=0.55]{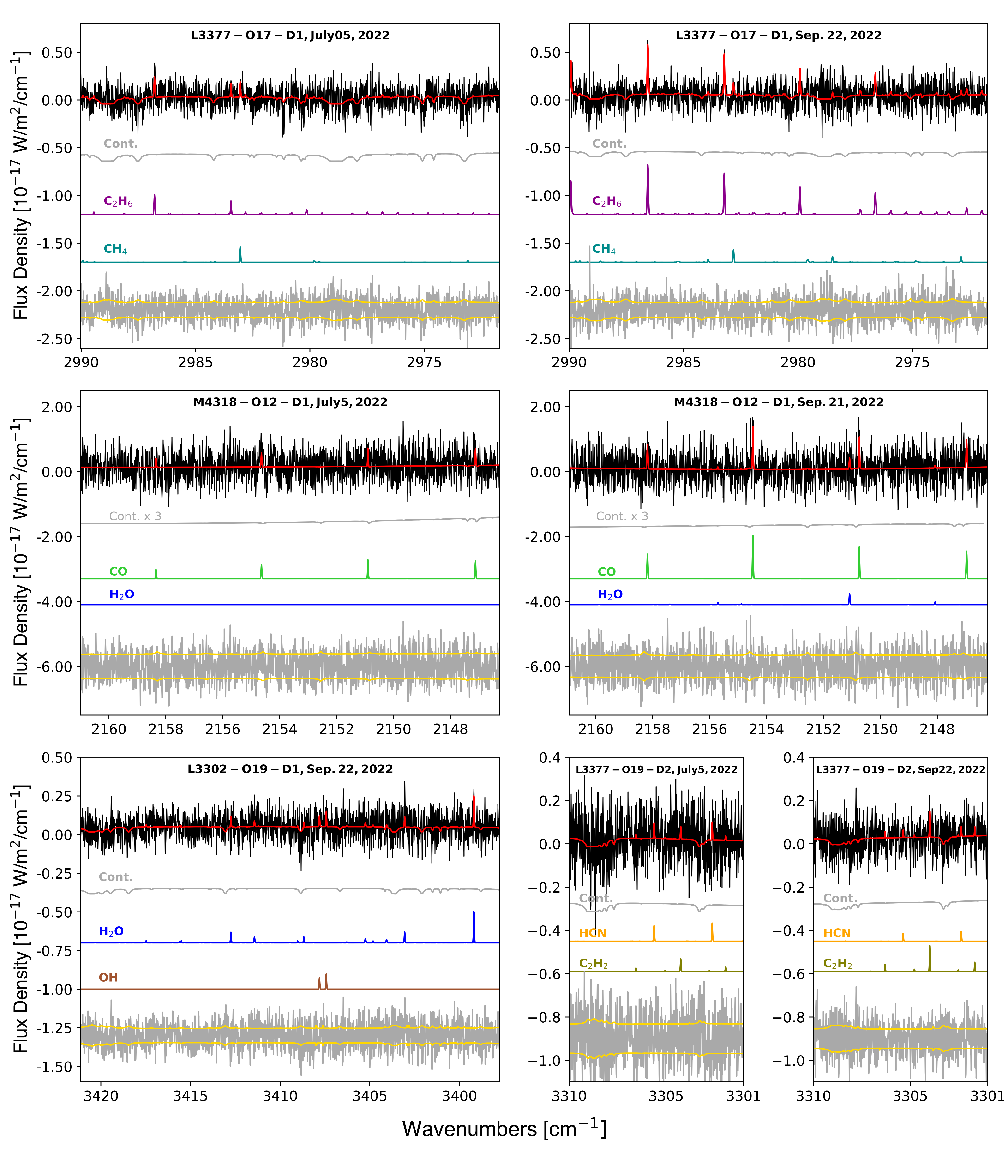}
	\caption{Spectra of comet C/2017 K2 acquired with CRIRES$^+$. In each plot, the top spectrum shows the observed data along with the total model (in red), while the gray spectrum below represents the modeled transmittance. Molecular models used to derive the production rates are displayed in different colors and labeled accordingly. Residuals are shown at the bottom in gray, with yellow lines indicating the $\pm$1$\sigma$ uncertainties. Setting, detector number, and observing date are shown on top of each plot.} 
	\label{fig:CRIRES_spectra}
\end{figure*}

\end{appendix}
\end{document}